\documentclass[reprint, aps, prd, letterpaper, noshowpacs, amsmath, %
amssymb, amsfonts, nofootinbib, floatfix, twoside]{revtex4-1}
\pdfoutput=1

\usepackage[T3,T1]{fontenc}
\usepackage{mathrsfs} 
\usepackage{bm} 
\makeatletter
\@ifclassloaded{beamer}
  { 
    \typeout{UsePackages: Detected beamer}
    \usepackage{tgheros}
    
  }
  { 
    \typeout{UsePackages: Did not detect beamer}
    \ifx\asybeamer\undefined 
    \typeout{UsePackages: Detected article}
    \usepackage[varg]{txfonts} 
    \usepackage{tgtermes} 
    \else 
    \typeout{Fonts: Detected asy for beamer}
    \usepackage{tgheros}
    
    \fi
  }
\usepackage{microtype} 
\def\MT@register@subst@font{
  \MT@exp@one@n\MT@in@clist\font@name\MT@font@list
  \ifMT@inlist@\else\xdef\MT@font@list{\MT@font@list\font@name,}\fi}
\makeatother

\usepackage{graphicx} %
\usepackage[x11names,svgnames,rgb]{xcolor} %
\usepackage{xspace}
\usepackage{braket}
\usepackage{accents}
\usepackage{siunitx}
\usepackage{mathtools}
\DeclareSymbolFontAlphabet{\mathrm}{operators}

\definecolor{CiteColor}{rgb}{0.18039, 0.18824, 0.57255}
\definecolor{UrlColor} {rgb}{0.741, 0.173, 0.000}
\definecolor{DarkUrlColor} {rgb}{0.500, 0.110, 0.000}
\definecolor{LinkColor}{rgb}{0.25098, 0.47843, 0.04706}

\makeatletter %
\newcommand{\ShowFont}{%
  \typeout{The main font is \f@encoding \space \f@family \space %
    \f@series \space \f@shape \space at \f@size pt.}%
  \typeout{The math font sizes are \tf@size pt (main), \sf@size pt %
    (script), and \ssf@size pt (scriptscript).}%
  \typeout{The linewidth is \the\linewidth}} %
\makeatother %

\usepackage{xargs}

\makeatletter

\def\@seccntformat#1{\csname the#1\endcsname.~}%

\def\section{%
  \@startsection {section}
  {1} {\z@} {0.55cm \@plus1ex \@minus .02ex}%
    {0.225cm} { \normalfont\bfseries \centering}%
}%
\def\subsection{%
  \@startsection {subsection}
  {2} {\z@ } {0.45cm \@plus 0.8ex \@minus 0.2ex}%
  {0.1125cm}{\normalfont \bfseries \centering }}
\def\subsubsection{%
  \@startsection {subsubsection}
  {3} {\z@ } {0.4cm \@plus 0.6ex \@minus 0.1ex}%
  {0.075cm}{\normalfont \it \centering }}

\newcommand{\surnames}[1]{\def\@surnamelist{#1}\relax}
\surnames{\ }
\ifx \@shorttitle \@empty
\else
  \def\@oddhead{\small \MakeUppercase{\@shorttitle} \hfill}
\fi
\ifx \@surnamelist \@empty
\else
  \def\@evenhead{\small \MakeUppercase{\@surnamelist} \hfill}
\fi
\def\@oddfoot{\reset@font\hfil\thepage\hfil}%
\def\@evenfoot{\reset@font\hfil\thepage\hfil}%
\g@addto@macro\maketitle{\global\@specialpagetrue\gdef\@specialstyle{plain}}

\usepackage[colorlinks, plainpages=false,
hyperfigures=true]{hyperref}
\hypersetup{linkcolor=LinkColor}
\hypersetup{citecolor=CiteColor}
\hypersetup{urlcolor=UrlColor}
\hypersetup{setpagesize=false}
\hypersetup{pdfborder=0 0 0}

\makeatother

\let\Originalcdefinition\c
\let\Originalddefinition\d
\let\Originaledefinition\e
\let\Originalidefinition\i
\DeclareMathOperator{\adjoint}{ad}
\makeatletter
\newcommand{\etal}{\textit{et~al}\@ifnextchar{\relax}{.\relax}{\ifx\@let@token.\else\ifx\@let@token~.\else.\@\xspace\fi\fi}}
\makeatother

\newcommand{\Cornell}{\affiliation{Cornell Center for Astrophysics and
    Planetary Science, Cornell University, Ithaca, New York 14853,
    USA}} %

\begin{document}


\graphicspath{%
  {Plots/}%
}

\title[The integration of angular velocity] {The integration of
  angular velocity}

\makeatletter \@booleantrue\frontmatterverbose@sw \makeatother

\surnames{Boyle}
\author{Michael Boyle} \Cornell

\date{\today}

\begin{abstract}
  A common problem in physics and engineering is determination of the
  orientation of an object given its angular velocity.  When the
  direction of the angular velocity changes in time, this is a
  nontrivial problem involving coupled differential equations.
  Several possible approaches are examined, along with various
  improvements over previous efforts.  These are then evaluated
  numerically by comparison to a complicated but analytically known
  rotation that is motivated by the important astrophysical problem of
  precessing black-hole binaries.  It is shown that a straightforward
  solution directly using quaternions is most efficient and accurate,
  and that the norm of the quaternion is irrelevant.  Integration of
  the generator of the rotation can also be made roughly as efficient
  as integration of the rotation.  Both methods will typically be
  twice as efficient naive vector- or matrix-based methods.
  Implementation by means of standard general-purpose numerical
  integrators is stable and efficient, so that such problems can be
  readily solved as part of a larger system of differential equations.
  Possible generalization to integration in other Lie groups is also
  discussed.
\end{abstract}



\maketitle

\section{Introduction}
\label{sec:Introduction}

In the study of kinematics, we sometimes know the angular velocity of
an object without knowing the actual orientation of that object.  The
angular velocity may be known from analysis of torques in dynamical
problems, or from data output by instruments in engineering
applications.  When the object is rotating in some complicated way,
finding the orientation from the angular velocity can be difficult; it
is neither conceptually obvious, nor computationally trivial.  This
paper presents several approaches that can be used to find the
orientation from a knowledge of the angular velocity.

Problems of this type arise in various situations.  In physics, we may
be able to calculate the torque exerted on an object by fluid drag,
gravitational or electrodynamics effects, or any of various forces.
If the moment-of-inertia tensor is known, the angular acceleration may
be calculated, from which a trivial integration will provide the
angular velocity.  The final step of deriving the orientation from the
angular velocity is less trivial.  Of course, this may be just one
part of a much more complicated whole, so that we end up with a
coupled system of differential equations, which must be solved
simultaneously.

The motivating example for this paper falls into the latter category.
The system is an astrophysical binary consisting of a pair of compact
objects such as neutron stars or black holes.  As the compact objects
orbit each other, they give off energy in the form of gravitational
waves and spiral in toward one another until they merge in one last
enormous burst of energy.  But the orbital dynamics can be quite
complicated.  The recent direct detection of gravitational
waves~\cite{GW150914} generated by a binary black-hole system has
introduced a new era in astronomy.  To capitalize on this new window
onto the universe, we need to be able to analyze the dynamics of these
systems very precisely and with maximum efficiency.  In particular,
when the black holes have significant spin angular momenta misaligned
with their orbital angular momentum, the binary will undergo
complicated precession and nutation as it orbits, which directly
affects the gravitational-wave signals given off.  The underlying
purpose of this paper is to enable accurate and efficient modeling of
precessing binaries.

But the applicability of this problem ranges far beyond
gravitational-wave astronomy.  The same mathematics find application
in mundane settings for engineering.  For example, the micromechanical
``gyroscopes'' ubiquitous in smartphones and other consumer
electronics provide the device's angular acceleration, but cannot
provide the instantaneous orientation due to low-frequency
noise.\footnote{Such an object is not a gyroscope in the traditional
  sense of the word, in that it has no spinning member.  Rather, it
  has what is essentially a spring-mounted inertial mass, and changes
  in the relative orientation of this mass are measured.  The coupling
  between the inertial mass and the surrounding device accounts for
  the loss of information at low frequencies.} The angular velocity
must therefore be integrated to arrive at the orientation.  Though the
principle may be the same, the noise and discrete nature of the
angular velocity suggest that low-order methods would be preferable
for these devices.  This, of course, is a particular example in the
broader field of strapdown inertial navigation
systems~\cite{woodman2007, miller1983, candy2012, wu2012}.  In a
completely different setting, Simo~\cite{simo1985} pointed out that
uniform beams undergoing twisting and bending obey equations formally
identical to those for angular velocity, where time is replaced by arc
length along the beam and the angular velocity is replaced by the rate
of twisting per unit length.  This particular problem has even been
analyzed using the methods of geometric algebra.  Such problems have
been analyzed previously in the literature~\cite{woodman2007,
  miller1983, candy2012, wu2012, mcrobie1999, ignagni1990,
  ignagni1990-a, ignagni1996}, but those approaches are designed very
specifically for pure rotational (and possibly translational) problems
without any other evolved quantities, and are frequently tailored even
further to approximate particular subclasses of motion.  They are
therefore not suitable when high accuracy is needed for long
integrations involving complicated rotations and other coupled
quantities evolving with multiple timescales.  In particular, much
additional work would be needed to develop adaptive integrators based
on these methods, and to incorporate them into general-purpose
integrators.

More broadly, techniques for integration in general Lie groups have
been suggested by various authors, including Magnus~\cite{magnus1954,
  hairer2006} and Munthe-Kaas~\cite{munthe-kaas1999, hairer2006}, who
modified the equations to evolve slightly different quantities so that
the solution would remain within the Lie manifold.  These approaches
are also discussed in Sec.~\ref{sec:note-about-munthe}, in comparison
to one of the approaches used in this paper.  On a more practical
level, numerical integration schemes have been proposed specifically
for Lie groups.  Crouch and Grossman~\cite{crouch1993, hairer2006}
suggested replacing the additive update method of standard numerical
integration schemes with a multiplicative method more suited to Lie
groups.  Those general techniques have evidently filtered back down to
influence the more theoretical approaches to problems involving the
familiar simple Lie groups.  Bottasso and Borri~\cite{bottasso1998}
developed an algorithm that appears to be essentially the
Crouch-Grossman technique specialized to the rotation group, while
simultaneously treating displacements.  Numerous groups have devised
other specialized low-order numerical algorithms for such
integrations~\cite{simo1991, lin1995, walton1993, munjiza2003,
  Bogfjellmo2014}, as well as several adapted to closely related
problems in Lagrangian and Hamiltonian dynamics~\cite{shivarama2004,
  kane2000, lew2004}.  Candy and Lasenby~\cite{Candy2011} outlined
integrations of rotations using quaternions, and simultaneous
treatment of rotations and translations using their natural
generalizations in the conformal representation of three-geometry.
Two studies compared several of these approaches using simple example
rotations---one by Johnson, Williams, and Cook~\cite{johnson2008}; the
other by Zupan and Saje~\cite{Zupan2011}.  Unfortunately, both focused
on low-order integration methods and achieved very large errors.  More
recently, Zupan and Zupan~\cite{zupan2014} and Treven and
Saje~\cite{treven2015} refined the earlier techniques to achieve much
improved stability and accuracy.  However, the goal of those works was
to find integrators specialized to the problem of rotation.  Even in
these more theoretical studies, integration of angular velocity has
apparently not been successfully tested using general-purpose adaptive
numerical integrators and physically motivated rotational examples
with coupled variables.

<The preceding shows that there are numerous possible ways of
integrating angular velocity---some being fairly subtle variations of
others.  Of course, in numerical analysis, minor differences can have
enormous effects on accuracy and efficiency.  For this paper, however,
we will not be particularly concerned with finer details of the
numerical integrators.  The reason for this is twofold.  First, many
systems involve integration of the angular velocity as just one part
of a much large set of equations, other members of which may not be
treatable with specialized methods, leaving us wanting to apply a
general-purpose integration routine.  In fact, we will find that two
such routines are capable of very accurate and efficient integration.
Second, and more importantly, substantial insight can be gained simply
by considering the more basic issue of exactly which set of equations
should be solved.  General considerations along these lines will
suggest the optimal quantities to be evolved.

When integrating angular velocity, the orientation we wish to
calculate is defined as a vector basis that is fixed in the rotating
system.  We call this the ``body'' frame---though there may not be a
clearly defined body involved, as in the case of compact binaries.
The body frame is defined with respect to an ``inertial'' frame that
is fixed in space.  The most obvious way to define the orientation,
then, is to simply express the body frame in the basis of the inertial
frame or vice versa.  Equivalently, we can define the orientation as
the rotation needed to transform the inertial frame into the body
frame.  Given the wide variety of ways that have been invented to
describe rotations, there are similarly many ways to integrate the
angular velocity to find the orientation.

Three possible methods for integrating angular velocity will be
examined in the following.  First is direct integration of the basis
vectors of the rotating frame, presented in
Sec.~\ref{sec:integr-frame-vectors}.  This is precisely equivalent to
integration of the rotation matrix if three orthonormal basis vectors
are used.  We can, however, improve the accuracy and efficiency of
this system by integrating just two of the basis vectors.  Of course,
this incorporates redundant information, which can---in principle---be
removed by constraint damping.  It turns out, however, that constraint
damping simply stiffens the system, making it far less efficient to
integrate.  In any case, it appears that the vector approach will
always be slower than the alternatives.

The second method examined will use the quaternion representing the
rotation, or ``rotor'', which can be obtained directly as described in
Sec.~\ref{sec:integr-spinors}.  There is one additional degree of
freedom in this representation, which can be considered the freedom to
renormalize the four components.  However, this freedom can easily be
made redundant simply by applying the quaternion in such a way that
the normalization cancels out; as long as the norm is nonzero, it is
irrelevant and there are effectively only three degrees of freedom in
the rotation represented by the quaternion.

Finally, we can also integrate the \emph{generator} of the rotation,
which inherently has just three degrees of freedom.  Using quaternion
methods again, this can be done without resorting to matrices and
their cumbersome methods of exponentiation.  There are reasons to
expect that generators could provide the most accurate and efficient
integration.  It turns out that a naive implementation will actually
be slower than even the vector implementation because the generator is
sometimes very sensitive to the rotation, so that it will change very
sharply.  This rapid behavior will force the integrator to take
smaller steps, slowing the integration considerably.  However, we can
impose a simple algebraic condition on the generator that avoids those
sharp features, making the generator approach robust, and efficient
enough to be competitive with even the rotor implementation.  The
derivation of this approach given in Sec.~\ref{sec:integr-bivectors}
will be more general than strictly necessary because the technique may
be interesting for application to integration in other Lie groups, as
discussed in Sec.~\ref{sec:note-about-munthe}.

In Sec.~\ref{sec:numerical-examples} these methods are then applied to
the problem that motivated this paper.  We first construct a framework
that can be used to devise very complicated rotations that can
nonetheless be expressed, along with their time derivatives, in closed
form.  It is then shown how this can be used to evaluate the accuracy
of the integration methods, including the definition of a useful and
complete measure of the error.  This framework is then applied to
construct a frame undergoing motion that corresponds very closely to
the motion experienced by a precessing black-hole binary.  This system
simply mimics several features of the binary---including fast rotation
about a main axis, slower precession around a gradually widening
precession cone, and small but fast nutation---but is given by an
analytical rotation formula, so that we can compare the numerical
results to the exact solution.  The particular rates of these
rotations are chosen to approximate the equivalent rates in a binary
that is very close to merger (the most dynamical stage of the binary's
life).  But the integration is carried out substantially longer than
is required by gravitational-wave data analysis, to provide a more
stringent test.  The result is excellent performance by the rotor
method and by the modified generator method, each approaching the
limits of numerical precision.  The vector method is roughly twice as
slow, and it is argued that this will be a very common feature, so
that we can ignore the vector approach, along with the associated
matrix approach.

While the simulated binary is a very particular example, it is also a
very rigorous one, incorporating large fast motions along with smaller
and slower motions.  This suggests that we should expect these results
to apply much more generally to other rotating systems.  Indeed,
several very different (though less realistic) additional examples
shown in the Appendix bear out these conclusions.  In particular,
while the generator method is certainly competitive, the rotor method
is likely to be slightly more efficient in many cases and possibly
more robust.  Thus, we can propose the rotor method as the general
choice for integrating rotations in three dimensions, while the
reasonable effectiveness of the generator suggests that it may be a
useful approach in more general Lie groups.

\section{Integration of frame vectors}
\label{sec:integr-frame-vectors}

Perhaps the most obvious approach to integrating the angular velocity
to arrive at a frame consists of simply integrating the frame vectors.
Assume the rotating system has a set of frame vectors
$\bm{f}_{i}$, which are stationary in the rotating system.  By
definition of the angular velocity, with respect to the inertial
system we have
\begin{equation}
  \label{eq:integr-frame-vectors}
  \frac{\ensuremath{\mathrm{d}} \bm{f}_{i}}{\ensuremath{\mathrm{d}} t} = \bm{\omega} \times
  \bm{f}_{i}.
\end{equation}
If the vectors $\bm{f}_{i}$ are unit vectors corresponding to the
$x$, $y$, and $z$ directions, this is precisely equivalent to
integrating the rotation matrix.

However, there is clearly some redundancy in this naive approach,
because we are integrating nine variables representing the three
components of each of these three vectors; on the other hand we know
that a rotation is determined by just three parameters.  Our naive
description is redundant because the equations ignore the fact that
the frame vectors have unit magnitude, and ignore their mutual
orthogonality.  We can reduce the redundancy by eliminating one of the
vectors: evolve only two of the vectors (say $\bm{f}_{0}$ and
$\bm{f}_{1}$), and deduce the third by taking their cross product
($\bm{f}_{2} = \bm{f}_{0} \times \bm{f}_{1}$).  It would
be possible to remove further degrees of freedom, for example by
computing only two of the components of $\bm{f}_{0}$, and
computing the third using the normalization of the vector.  But this
would leave it determined only up to a sign, so we would need
additional logic to choose that sign.  The additional complications
are likely not worth the trouble.

During a numerical integration, any violation of the constraints due
to finite precision could grow.  In that case, there are additional
methods of enforcing constraints---the standard methods being damping
and projection~\cite{hairer2006}.  Constraint projection involves
simply replacing the solution after each time step with the nearest
solution that satisfies the constraints.  In this case, one could use
a standard Gram-Schmidt orthonormalization procedure.  Constraint
damping involves the addition of terms to the differential equations
that drive the solution back toward the constraint surface.  For
example, to enforce the constraint that elements of our frame have
unit norm, we could modify the equation above to read
\begin{equation}
  \label{eq:vector-norm-damping}
  \frac{\ensuremath{\mathrm{d}} \bm{f}_{i}}{\ensuremath{\mathrm{d}} t} = \bm{\omega} \times
  \bm{f}_{i} + \lambda_{i} \left(1 - \bm{f}_{i} \cdot
    \bm{f}_{i}\right)\, \bm{f}_{i},
\end{equation}
where $\lambda_{i}$ is some constant that sets the timescale on which
constraint violations are damped as $1/\lambda_{i}$.  When the norm of
the vector is $1$, this does not modify the evolution in any way; when
it is different, this simply forces the vector back to the nearest
constrained value.  It must be noted, however, that this makes the
equations stiff, which typically results in less efficient
integration.  We could also add terms to damp violations of the
orthogonality constraint, as in
\begin{equation}
  \label{eq:vector-orthogonality-damping}
  \frac{\ensuremath{\mathrm{d}} \bm{f}_{1}}{\ensuremath{\mathrm{d}} t} = \bm{\omega} \times
  \bm{f}_{1} + \mu\, \left( \frac{\bm{f}_{0} \times
      (\bm{f}_{1}\times \bm{f}_{0})} {\left\lvert{\bm{f}_{0}
        \times (\bm{f}_{1}\times \bm{f}_{0})}\right\rvert} -
    \bm{f}_{1} \right),
\end{equation}
for some constant $\mu$.  When $\bm{f}_{1}$ has unit norm and is
orthogonal to $\bm{f}_{0}$, the terms in parentheses cancel out,
so there is no change in the evolution; otherwise, that factor drives
$\bm{f}_{1}$ toward perpendicularity in the
$\bm{f}_{0}$-$\bm{f}_{1}$ plane.  Again, however, these
elaborate alterations can be expected to increase the complexity and
decrease the robustness of the solution, so it is not clear how useful
they can be.

\section{Integration of rotors}
\label{sec:integr-spinors}

Though integration of the frame vectors presents a very clear and
effective way of integrating angular velocity, the redundant
information is unappealing from an aesthetic standpoint.  Furthermore,
in numerical applications, we can expect that the redundancy might
reduce the accuracy and efficiency of integration.  Thus, it might be
preferable to find an alternative method of integrating.  Ultimately,
we are interested in computing a rotation that corresponds to the
given angular velocity, so we might wonder whether it is possible to
\emph{directly} find that rotation.  That is, we can search for the
operator performing the rotation, rather than the vectors being
rotated.  While rotation matrices are probably the most familiar way
of approaching rotations, we noted above that this would be equivalent
to integrating three orthogonal frame vectors, having nine degrees of
freedom to represent a three-dimensional problem.  A more elegant and
modern approach---yet simpler and easier to use, once they are
understood---is found in quaternions.  Though the early history was
tragically muddled by the misguided idea that vectors and quaternions
were incompatible~\cite{Crowe1985}, they are now seen as two aspects
of a more powerful unifying theory: geometric
algebra~\cite{Hestenes1987, Doran2010}, in which both quaternions and
the more familiar vector algebra arise as specializations in three
dimensions.  Because of their advantageous numerical and conceptual
properties, quaternions have found successful application in various
fields including computer graphics, robotics, molecular dynamics,
celestial mechanics, and orbital dynamics~\cite{Vold1993, Doran2010}.
Here, we find that they also give rise to a simple method for
integrating angular velocity.

Quaternions have a scalar part and a ``vector'' part,\footnote{In
  fact, geometric algebra makes clear that the ``vector'' part would
  be more coherently called a ``bivector'' part~\cite{Doran2010},
  where rather than a vector representing the axis of a rotation the
  bivector represents the plane in which the rotation takes place.
  This generalizes perfectly to vector spaces of any dimension and
  signature.  Coincidentally, a bivector in three dimensions has three
  components, just like a vector.  Misunderstanding of this accidental
  equality was the origin of the acrimonious debate between vector
  partisans and quaternion partisans in the late nineteenth
  century~\cite{Crowe1985}.  Even in three dimensions clarifying the
  distinction can lead to deeper understanding of the geometry, but we
  will use more standard terminology in this paper.} so that vectors
may be considered quaternions with scalar part $0$, and quaternions
with scalar part $0$ may be considered vectors equivalently.  At this
level, quaternions may be considered precisely four-dimensional
vectors.  However, there is a crucial additional structure provided
with them: they can be multiplied together.  The details of this
multiplication are available in any standard reference---perhaps the
best being Ref.~\cite{Doran2010}---but two salient features are that
it is associative but not commutative.  For our purposes, the most
interesting application of multiplication is the rotation of a vector.
Given a vector $\bm{v}$ and a nonzero quaternion $\mathbf{\MakeUppercase{R}}$, we can
construct a new vector\footnote{For numerical applications, rather
  than using quaternion multiplication directly, it is roughly twice
  as efficient to implement this equation as
  ${\bm{v}}' = \bm{v} + 2\, \bm{r} \times (s\, \bm{v}
  + \bm{r} \times \bm{v})/m$.  Here $s$ and $\bm{r}$ are
  the scalar and vector parts of $\mathbf{\MakeUppercase{R}}$, and $m$ is the sum of the
  squares of the components of $\mathbf{\MakeUppercase{R}}$.}
\begin{equation}
  \label{eq:rotor-rotation}
  {\bm{v}}' = \mathbf{\MakeUppercase{R}}\, \bm{v}\, \mathbf{\MakeUppercase{R}}^{-1}.
\end{equation}
It turns out that ${\bm{v}}'$ is simply the rotation of
$\bm{v}$ about the axis given by the vector part of $\mathbf{\MakeUppercase{R}}$, and the
angle of that rotation is related to the ratio of the scalar part to
the magnitude of the vector part.  Because the quaternion effects the
rotation of vectors, Clifford named this object a
``rotor''~\cite{Clifford1878}.  Then, rather than analyzing the
evolution of several vectors being rotated, we will analyze the
evolution of the single rotor effecting this rotation.  It is true
that a quaternion has four elements, and hence one more degree of
freedom than the space of rotations being modeled; nonetheless, this
is a substantial improvement over the six to nine components of the
vector/matrix approach.

Before we see how this gives rise to an evolution equation, a brief
note is in order.  Clifford actually imposed another requirement on
what he would call a rotor, which is also usually imposed elsewhere in
the literature.  That is: a rotor should have unit magnitude, meaning
that the sum of the squares of the scalar and vector components should
equal \num{1}.  Then, rather than Eq.~\eqref{eq:rotor-rotation},
Clifford and others generally use
${\bm{v}}' = \mathbf{\MakeUppercase{R}}\, \bm{v}\, \bar{\mathbf{\MakeUppercase{R}}}$, where $\bar{\mathbf{\MakeUppercase{R}}}$ is the
conjugate which simply reverses the sign of the vector part of $\mathbf{\MakeUppercase{R}}$.
When the magnitude is \num{1}, these are precisely equivalent.
However, due to numerical error, this constraint may be
violated.\footnote{It is also worth noting that there are cases in
  which the alternative form of Eq.~\eqref{eq:rotor-rotation} using
  the conjugate in place of the inverse may be used to describe
  changes in a vector that should \emph{not} preserve its magnitude,
  using quaternions that intentionally have non-unit magnitude.  For
  example, the quaternion may be used to model eccentric orbits---in
  which case the magnitude should change as the orbiting body traces
  out the ellipse---leading to enormous
  simplifications~\cite{hestenes1983, Doran2010}.}  If we were to use
the conjugate instead of the inverse, ${\bm{v}}'$ would have a
different magnitude than $\bm{v}$, and so would no longer be a
rotation.  Using the inverse instead makes the magnitude
irrelevant---as long as it is nonzero so that an inverse actually
exists.  We are abusing language slightly in calling these objects
rotors (the more standard term being ``spinors''), but the intent is
really identical.

Now, to derive an evolution equation, we assume that $\bm{v}$ is
some constant vector, perhaps representing an initial value, while
${\bm{v}}'(t)$ is evolving in the inertial frame.  This will
correspond via Eq.~\eqref{eq:rotor-rotation} to some rotor $\mathbf{\MakeUppercase{R}}(t)$,
whose evolution we will now relate to $\bm{\omega}$ (dropping the
arguments $t$ for simplicity).  First, we can use the product rule to
differentiate $\mathbf{\MakeUppercase{R}}\, \mathbf{\MakeUppercase{R}}^{-1} = 1$ and find
\begin{equation}
  \label{eq:differentiate-unit-condition}
  \frac{\ensuremath{\mathrm{d}}}{\ensuremath{\mathrm{d}} t} \mathbf{\MakeUppercase{R}}^{-1} = -\mathbf{\MakeUppercase{R}}^{-1}\, \frac{\ensuremath{\mathrm{d}}
    \mathbf{\MakeUppercase{R}}} {\ensuremath{\mathrm{d}} t}\, \mathbf{\MakeUppercase{R}}^{-1}.
\end{equation}
Using this result, it is not hard to show that
\begin{equation}
  \label{eq:derivative-of-vector}
  \frac{\ensuremath{\mathrm{d}} \bm{v}'}{\ensuremath{\mathrm{d}} t}
  = \frac{\ensuremath{\mathrm{d}}}{\ensuremath{\mathrm{d}} t} \left( \mathbf{\MakeUppercase{R}}\, \bm{v}\,
    \mathbf{\MakeUppercase{R}}^{-1} \right)
  = \left[\frac{\ensuremath{\mathrm{d}} \mathbf{\MakeUppercase{R}}}{\ensuremath{\mathrm{d}} t} \mathbf{\MakeUppercase{R}}^{-1},
    \bm{v}'\right],
\end{equation}
where the brackets in the last expression denote the usual commutator.
Similar calculations show that the term $\frac{\ensuremath{\mathrm{d}}\mathbf{\MakeUppercase{R}}}{\ensuremath{\mathrm{d}} t}\, \mathbf{\MakeUppercase{R}}^{-1}$
is a pure vector if and only if the quantity $\mathbf{\MakeUppercase{R}}\, \bar{\mathbf{\MakeUppercase{R}}}$ is
constant in time (though it need not equal 1).  In that case, a simple
exercise using the definition of quaternion multiplication shows that
we can rewrite this last expression as
\begin{equation}
  \label{eq:derivative-of-vector-cross}
  \frac{\ensuremath{\mathrm{d}} \bm{v}'}{\ensuremath{\mathrm{d}} t}
  = \left( 2 \frac{\ensuremath{\mathrm{d}} \mathbf{\MakeUppercase{R}}}{\ensuremath{\mathrm{d}} t} \mathbf{\MakeUppercase{R}}^{-1}
  \right) \times \bm{v}',
\end{equation}
where the symbol $\times$ is just the usual vector cross product.
Now, recalling the standard angular-velocity formula
\begin{equation}
  \label{eq:angular-velocity-as-derivative}
  \frac{\ensuremath{\mathrm{d}} \bm{v}'}{\ensuremath{\mathrm{d}} t} = \bm{\omega} \times \bm{v}',
\end{equation}
and noting that these results hold for arbitrary $\bm{v}$, we see
that
\begin{equation}
  \label{eq:angular-velocity}
  \bm{\omega} = 2 \, \frac{\ensuremath{\mathrm{d}} \mathbf{\MakeUppercase{R}}}{\ensuremath{\mathrm{d}} t}\,
  \mathbf{\MakeUppercase{R}}^{-1}.
\end{equation}
As promised, this relates the rotor $\mathbf{\MakeUppercase{R}}$ to the angular velocity
$\bm{\omega}$.  In fact, we can turn this into a first-order
differential equation:
\begin{equation}
  \label{eq:integrate-spinor}
  \frac{\ensuremath{\mathrm{d}} \mathbf{\MakeUppercase{R}}}{\ensuremath{\mathrm{d}} t} = \frac{1}{2} \bm{\omega}\,
  \mathbf{\MakeUppercase{R}}.
\end{equation}
As long as $\bm{\omega}$ is known as a function of time and
possibly of $\mathbf{\MakeUppercase{R}}$ (but not of $\ensuremath{\mathrm{d}} \mathbf{\MakeUppercase{R}} / \ensuremath{\mathrm{d}} t$), standard theorems on
ordinary differential equations apply, which show that given initial
data for $\mathbf{\MakeUppercase{R}}$, we can simply evolve the four components of this
equation to find $\mathbf{\MakeUppercase{R}}$ as a function of time.\footnote{If we simply
  \emph{assume} that $\mathbf{\MakeUppercase{R}}$ obeys an equation of the
  form~\eqref{eq:integrate-spinor}, for a general quaternion
  $s + \bm{\omega}$ with scalar part $s$, then
  $\ensuremath{\mathrm{d}} (\mathbf{\MakeUppercase{R}} \bar{\mathbf{\MakeUppercase{R}}}) / \ensuremath{\mathrm{d}} t = s\, \mathbf{\MakeUppercase{R}}\, \bar{\mathbf{\MakeUppercase{R}}} / 2$.  But since we
  impose Eq.~\eqref{eq:integrate-spinor} with $s=0$, we know that
  $\mathbf{\MakeUppercase{R}}\, \bar{\mathbf{\MakeUppercase{R}}}$ is constant in time.  This shows that the conversion
  of Eq.~\eqref{eq:derivative-of-vector} into
  Eq.~\eqref{eq:derivative-of-vector-cross} is, at least, consistent.}

We will see below that quaternions can be exponentiated readily.  One
might then expect that Eq.~\eqref{eq:integrate-spinor} could be
integrated using the exponential:
\begin{equation}
  \label{eq:integr-const-ang-vel}
  \mathbf{\MakeUppercase{R}}(t) = \exp \left[ \frac{1}{2}\, \int_{0}^{t}
    \bm{\omega}(t')\, \ensuremath{\mathrm{d}} t' \right] \mathbf{\MakeUppercase{R}}(0).
\end{equation}
This actually may be a valid solution to the differential equation,
but only when $\bm{\omega}(t_{1})$ commutes with
$\bm{\omega}(t_{2})$ for any pair of times $t_{1}$ and $t_{2}$,
and with $\mathbf{\MakeUppercase{R}}(0)$---that is, when the rotation all takes place about a
single axis.  The quaternion exponential is substantially more
complicated than the real and complex exponentials, because of the
noncommutativity of the quaternion product.  More generally, when the
direction of $\bm{\omega}$ varies in time, we do not have
commutativity, and we must therefore treat the problem as a coupled
system of ordinary differential equations for the four components of
$\mathbf{\MakeUppercase{R}}(t)$.

One important point to note about our result,
Eq.~\eqref{eq:integrate-spinor}, is that we derived it without any
constraints.  The only assumptions that went into the derivation were
differentiability of the various elements, and the existence of an
inverse of $\mathbf{\MakeUppercase{R}}$---which will be the case as long as $\mathbf{\MakeUppercase{R}} \neq 0$.  In
particular, we did not assume that $\mathbf{\MakeUppercase{R}}$ has unit magnitude.
Interestingly enough, Eq.~\eqref{eq:integrate-spinor} results when we
use the alternative form of Eq.~\eqref{eq:rotor-rotation} mentioned
above in which unit magnitude is assumed.  Thus, we could think of our
approach as a form of constraint projection.  In fact, the rotation
group $\mathrm{SO}(3)$ is topologically the same as $\mathbb{R}\mathrm{P}^{3}$, which is usually
described as $\mathbb{R}^{4}$ with the origin removed, subject to
identification of all points that are scalar multiples of each
other~\cite{Hatcher2001}.  In a sense, we are evolving in that larger
space, and making the identification later.  But it is clear that
Eq.~\eqref{eq:integrate-spinor} is the correct evolution equation,
\emph{even in the larger space}.  Many of the references cited in
Sec.~\ref{sec:Introduction} go to great lengths to enforce the unit
magnitude of the rotor; the argument here suggests that such effort is
wholly unnecessary.

\section{Integration of generators}
\label{sec:integr-bivectors}

The solution given by Eq.~\eqref{eq:integr-const-ang-vel} suggests
another possible approach to integrating angular velocity.  The
quantity in the exponential is called the generator of the rotation,
and is linear in the angular velocity for this solution.  It is true
that this solution is not valid for more general angular velocities,
in which the direction of the angular velocity changes.  But perhaps
the general solution may at least be \emph{dominated} by linear
behavior, which would suggest useful perturbative expansions and
effective numerical implementation.  This approach also finds
motivation in the formalism of Lie theory.  The rotation group is a
Lie group, and every Lie group is governed locally by its
corresponding Lie algebra; a path through the group is given by
integrating the differential motions given by elements of the algebra,
in complete generality for \emph{all} Lie groups.  The relation
between the group and the algebra is given---locally at least---by the
exponential function.  Thus, in this section, we investigate computing
the \emph{generator} of the rotation, rather than the rotation itself.
The derivation used here was first introduced in
Ref.~\cite{Boyle2013}, though the key equation,
Eq.~\eqref{eq:log_deriv_omega}, was derived much earlier using a very
different approach by Grassia~\cite{Grassia1998} and then by Candy and
Lasenby~\cite{Candy2011} using still another approach.  The present
derivation is more complicated, but has the advantage of generalizing
to other Lie groups.

To begin, we relate a rotor $\mathbf{\MakeUppercase{R}}$ to its generator $\mathbf{\MakeLowercase{r}}$
by $\mathbf{\MakeUppercase{R}} = \ensuremath{\mathrm{e}}^{\mathbf{\MakeLowercase{r}}}$.  Here the generator
$\mathbf{\MakeLowercase{r}}$ is a quaternion with zero scalar part, which is
sometimes called a pure vector, and exponentiation is given by the
usual series expression incorporating integer powers of the generator.
For rotations, this vector $\mathbf{\MakeLowercase{r}}$ has the interpretation
of being the axis about which the rotation takes place, and its
magnitude is half the angle of that rotation.  In any case, Lie theory
then provides us with a formula~\cite{*[][{.  See Sec.~1.5.}]
  Duistermaat2000} for the derivative of $\mathbf{\MakeUppercase{R}}$ in terms of
$\mathbf{\MakeLowercase{r}}$ and its derivative:
\begin{equation}
  \label{eq:rotor_derivative_Lie_group}
  \frac{\ensuremath{\mathrm{d}} \mathbf{\MakeUppercase{R}}}{\ensuremath{\mathrm{d}} t}
  = \frac{\ensuremath{\mathrm{d}} \ensuremath{\mathrm{e}}^{\mathbf{\MakeLowercase{r}}}}{\ensuremath{\mathrm{d}} t}
  = \int_{0}^{1} \ensuremath{\mathrm{e}}^{s\, \mathbf{\MakeLowercase{r}}}\, \frac{\ensuremath{\mathrm{d}}
    \mathbf{\MakeLowercase{r}}} {\ensuremath{\mathrm{d}} t} \, \ensuremath{\mathrm{e}}^{(1-s)\,\mathbf{\MakeLowercase{r}}}\, \ensuremath{\mathrm{d}} s.
\end{equation}
Multiplying this equation on the right by
$\mathbf{\MakeUppercase{R}}^{-1} = \ensuremath{\mathrm{e}}^{-\mathbf{\MakeLowercase{r}}}$, it is clear that we will obtain a
formula relating the angular velocity~\eqref{eq:angular-velocity} to
the generator of the rotation and its time-derivative.  To put this
into a more useful form, however, we need to evaluate the integral and
simplify.

We can write a standard formula~\cite{Miller1972} as
\begin{equation}
  \label{eq:conjugate_adjoint}
  \ensuremath{\mathrm{e}}^{\mathbf{\MakeLowercase{p}}}\, \mathbf{\MakeLowercase{q}}\, \ensuremath{\mathrm{e}}^{-\mathbf{\MakeLowercase{p}}}
  = \sum_{k=0}^{\infty} \frac{1}{k!} \adjoint_{\mathbf{\MakeLowercase{p}}}^{k} \mathbf{\MakeLowercase{q}},
\end{equation}
where the adjoint function is defined recursively by
\begin{subequations}
  \label{eq:define_adjoint}
  \begin{align}
    \adjoint_{\mathbf{\MakeLowercase{p}}}^{0} \mathbf{\MakeLowercase{q}}
    &= \mathbf{\MakeLowercase{q}}, %
    \\
    \adjoint_{\mathbf{\MakeLowercase{p}}}^{1} \mathbf{\MakeLowercase{q}}
    &= [\mathbf{\MakeLowercase{p}}, \mathbf{\MakeLowercase{q}}],
    \\
    \adjoint_{\mathbf{\MakeLowercase{p}}}^{k+1} \mathbf{\MakeLowercase{q}}
    &= \left[ \mathbf{\MakeLowercase{p}}, \adjoint_{\mathbf{\MakeLowercase{p}}}^{k} \mathbf{\MakeLowercase{q}} \right],
  \end{align}
\end{subequations}
and
$[\bm{a}, \bm{b}] = \bm{a} \bm{b} - \bm{b}
\bm{a}$ represents the Lie bracket.  We now have
\begin{equation}
  \label{eq:derivative_Lie_group}
  \bm{\omega} = 2\frac{\ensuremath{\mathrm{d}} \mathbf{\MakeUppercase{R}}}{\ensuremath{\mathrm{d}} t}\,
  \mathbf{\MakeUppercase{R}}^{-1} = 2\int_{0}^{1} \sum_{k=0}^{\infty} \frac{1}{k!}
  \adjoint_{s\, \mathbf{\MakeLowercase{r}}}^{k} \frac{\ensuremath{\mathrm{d}} \mathbf{\MakeLowercase{r}}} {\ensuremath{\mathrm{d}}
    t}\, \ensuremath{\mathrm{d}} s.
\end{equation}
Formally, at least, this fulfills our objective of relating
$\ensuremath{\mathrm{d}}\mathbf{\MakeLowercase{r}} / \ensuremath{\mathrm{d}} t$ to $\bm{\omega}$ and
$\mathbf{\MakeLowercase{r}}$.  It should be noted that, up to this point, the
derivation has been completely general\footnote{We assumed that it is
  always possible to find a generator satisfying
  $\mathbf{\MakeUppercase{R}} = \ensuremath{\mathrm{e}}^{\mathbf{\MakeLowercase{r}}}$ because the rotation group is
  connected, although we will see below that such a generator is not
  unique.  For more general Lie groups, only an element in the
  connected component of the identity can be written in this way, but
  arbitrary elements may be written as the product of such an
  exponential and some (componentwise-constant) element in the
  connected component of the result:
  $\mathbf{\MakeUppercase{R}} = \ensuremath{\mathrm{e}}^{\mathbf{\MakeLowercase{r}}}\, \mathbf{\MakeUppercase{R}}_{0}$.  The derivation of
  Eq.~\eqref{eq:derivative_Lie_group} follows in exactly the same
  way.} and applies to any Lie algebra.  This will be discussed
further in Sec.~\ref{sec:note-about-munthe}.

We can now specialize to $\mathfrak{so}(3)$, and use induction to prove
\begin{equation}
  \label{eq:adjoint-powers}
  \adjoint_{\mathbf{\MakeLowercase{p}}}^{k} \mathbf{\MakeLowercase{q}} =
  \begin{cases}
    \mathbf{\MakeLowercase{q}} & k=0; \\
    (-1)^{(k-1)/2}\, [\mathbf{\MakeLowercase{p}}, \mathbf{\MakeLowercase{q}}]\, (2\,
    \mathrm{\MakeLowercase{p}})^{k-1} & \text{$k>0$ odd}; \\
    (-1)^{(k-2)/2}\, [\mathbf{\MakeLowercase{p}}, [\mathbf{\MakeLowercase{p}},
    \mathbf{\MakeLowercase{q}}]]\, (2\, \mathrm{\MakeLowercase{p}})^{k-2} & \text{$k>0$
      even}.
  \end{cases}
\end{equation}
Plugging this expression into Eq.~\eqref{eq:derivative_Lie_group},
evaluating the sum, and integrating, we find
\begin{equation}
  \label{eq:omega_logarithms}
  \bm{\omega} = 2\, \dot{\mathbf{\MakeLowercase{r}}} + \frac{\sin^2
    \mathrm{\MakeLowercase{r}}} {\mathrm{\MakeLowercase{r}}^2}
  \Big[\mathbf{\MakeLowercase{r}}, \dot{\mathbf{\MakeLowercase{r}}}\Big] +
  \frac{\mathrm{\MakeLowercase{r}} - \sin \mathrm{\MakeLowercase{r}}\, \cos
    \mathrm{\MakeLowercase{r}}} {2 \mathrm{\MakeLowercase{r}}^3} \bigg[
  \mathbf{\MakeLowercase{r}}, \Big[ \mathbf{\MakeLowercase{r}}, \dot{\mathbf{\MakeLowercase{r}}}
  \Big] \bigg].
\end{equation}
For $\mathrm{\MakeLowercase{r}} = \pi$, we have $\mathbf{\MakeLowercase{r}} = -1$, and
so this reduces to $\dot{\mathbf{\MakeLowercase{r}}} = \bm{\omega}/2$.  For
smaller values of $\mathrm{\MakeLowercase{r}}$, we can solve this equation
for $\dot{\mathbf{\MakeLowercase{r}}}$.  Using the fact that in three
dimensions,
$[\bm{a}, \bm{b}] = 2\, \bm{a} \times \bm{b}$ (with
$\times$ representing the usual vector cross product), we can write
the solution as
\begin{equation}
  \label{eq:log_deriv_omega}
  \dot{\mathbf{\MakeLowercase{r}}} = \frac{1}{2} \bm{\omega}
  \times \mathbf{\MakeLowercase{r}} + \bm{\omega}\,
  \frac{\mathrm{\MakeLowercase{r}}\, \cot \mathrm{\MakeLowercase{r}}} {2} +
  \mathbf{\MakeLowercase{r}}\, \frac{\mathbf{\MakeLowercase{r}} \cdot \bm{\omega}}
  {2\, \mathrm{\MakeLowercase{r}}^{2}} ( 1 - \mathrm{\MakeLowercase{r}}\, \cot
  \mathrm{\MakeLowercase{r}} ).
\end{equation}
This differential equation for $\mathbf{\MakeLowercase{r}}$ can be used to
integrate the orientation of the system given $\bm{\omega}$ as a
function of $t$ and possibly $\mathbf{\MakeLowercase{r}}$, and so to find
$\mathbf{\MakeLowercase{r}}$ as a function of time.

One point to note here is that the magnitude of $\mathbf{\MakeLowercase{r}}$ can
be unbounded when integrating this equation.  In realistic
applications, this would happen very rarely, because it requires the
rotation to return to the identity rotation.  However, if the rotation
is restricted to a fixed axis, or if there is some other reason the
system should happen to return to the identity, this can occur.  And
it can cause problems in numerical integrations when the system
returns \emph{approximately} to the identity.  At these times, the
integration becomes very delicate, requiring high precision to retain
accuracy in the result.  We can avoid this condition, however, using
the fact that different values of the generator represent the same
rotation.  In particular,
\begin{equation}
  \label{eq:rotation_equivalence}
  \exp[\mathbf{\MakeLowercase{r}}] \quad \text{and} \quad \exp \left[
    \mathbf{\MakeLowercase{r}} + n\, \pi \frac{\mathbf{\MakeLowercase{r}}}
    {\mathrm{\MakeLowercase{r}}} \right]
\end{equation}
represent identical rotations for any integer $n$.\footnote{This is a
  generalization to quaternions of the more familiar result from
  complex algebra $\ensuremath{\mathrm{e}}^{z} = (-1)^{n}\, \ensuremath{\mathrm{e}}^{z + \pi\, i\, n}$.  The
  negative sign is irrelevant for our purposes because the rotor is
  used twice to rotate a vector, so the sign drops out.} Thus,
whenever $\mathrm{\MakeLowercase{r}} \geq \pi/2$, we can reset the value of
the generator according to
\begin{equation}
  \label{eq:generator_reset}
  \mathbf{\MakeLowercase{r}} \to \mathbf{\MakeLowercase{r}} - \pi \frac{\mathbf{\MakeLowercase{r}}}
  {\mathrm{\MakeLowercase{r}}}.
\end{equation}
When $\mathrm{\MakeLowercase{r}} = \pi/2$, this transformation is simply
$\mathbf{\MakeLowercase{r}} \to -\mathbf{\MakeLowercase{r}}$.  This means that rather than
rotating through $\pi$ about the given axis, we rotate by $\pi$ about
the axis in the opposite sense---which is an equivalent rotation.  For
$\mathrm{\MakeLowercase{r}} > \pi/2$ this reduces the magnitude of the
generator below $\pi/2$.  But in either case the derivative now
changes so that, rather than increasing back toward $\pi/2$, the
magnitude begins to decrease.  We will find below that this is a
crucial step in making integration of angular velocity by generators
an efficient approach.


\section{Integration in general Lie groups}
\label{sec:note-about-munthe}
Though not directly related to the purpose of this paper, the previous
section may be useful in very different situations, which we now take
a brief detour to discuss.  As noted below
Eq.~\eqref{eq:derivative_Lie_group}, the derivation of that expression
was entirely general, and applies to any Lie group.  The resulting
approach to integration is very similar in motivation to algorithms
described by Magnus~\cite{magnus1954, hairer2006} and
Munthe-Kaas~\cite{munthe-kaas1999, hairer2006} for integration in Lie
groups.  However, both assumed that expressions comparable to the sum
in Eq.~\eqref{eq:derivative_Lie_group} could not be calculated
explicitly, and would need to be truncated at some finite term
instead.  Higher powers of the adjoint simplify in our case using
properties of $\mathfrak{so}(3)$, which results in a recursive expression that
can be summed explicitly, avoiding truncation.

For more general Lie groups, we can expect a similar simplification
whenever there exists some $K$ such that for all $\mathbf{\MakeLowercase{p}}$
and $\mathbf{\MakeLowercase{q}}$, $\adjoint_{\mathbf{\MakeLowercase{p}}}^{K} \mathbf{\MakeLowercase{q}}$ can be written
as a linear combination of the $\adjoint_{\mathbf{\MakeLowercase{p}}}^{k} \mathbf{\MakeLowercase{q}}$ with
$k<K$.  This will always be true for nilpotent algebras; indeed, for
nilpotent algebras there exists a $K$ such that
$\adjoint_{\mathbf{\MakeLowercase{p}}}^{K} \mathbf{\MakeLowercase{q}} = 0$, which leaves us with a summation
over a finite series of terms.  However, ours is a much weaker
condition than nilpotency, and so will be true for a more general
class of Lie groups.

Other important problems for which such a simplification occurs
include those for which the evolution remains confined to a
three-dimensional subspace.  To see how, we need to use the form of
the Lie bracket as given by an antisymmetrized product between two
vectors:
$[\bm{a}, \bm{b}] = \bm{a}\, \bm{b} - \bm{b}\,
\bm{a}$.  In general, the products on the right-hand side of this
expression need not be defined; a Lie algebra only requires a
definition of the bracket.  However, there is a construction called
the universal enveloping algebra~\cite{Hall2015}, whereby every Lie
algebra can be expressed as a subspace of an associative algebra in
which such products \emph{are} defined.  We can simply regard the
original algebra as being embedded within the enveloping algebra---and
thus use the same symbols for notational simplicity.  Using the
associative product from the enveloping algebra, we have quite
generally\footnote{Again, this is proven using a simple induction,
  along with the definition of the adjoint,
  Eqs.~\eqref{eq:define_adjoint}.}
\begin{equation}
  \label{eq:power_of_adjoint}
  \adjoint_{\mathbf{\MakeLowercase{p}}}^{k} \mathbf{\MakeLowercase{q}} = \sum_{j=0}^{k} (-1)^{j+k}\,
  \binom{k}{j}\, \mathbf{\MakeLowercase{p}}^j\, \mathbf{\MakeLowercase{q}}\,
  \mathbf{\MakeLowercase{p}}^{k-j}.
\end{equation}
Whenever $\mathbf{\MakeLowercase{p}}^\ell$ is a scalar for some $\ell < k$, we
can pull such a factor out of the terms in this sum.  Then
$\adjoint_{\mathbf{\MakeLowercase{p}}}^{k} \mathbf{\MakeLowercase{q}}$ can be written as a linear combination
of $\adjoint_{\mathbf{\MakeLowercase{p}}}^{k-\ell} \mathbf{\MakeLowercase{q}}$ and lower-order terms.  Because
the adjoint is defined recursively, all higher-order terms will
similarly collapse, so that the sum in Eq.~\eqref{eq:omega_logarithms}
can be written using only a finite number of adjoints.  It may be
useful to consider these statements applied to the case of $\mathfrak{so}(3)$,
using quaternions.  The generator $\mathbf{\MakeLowercase{p}}$ is a ``pure
vector'' quaternion, and we know that
$\mathbf{\MakeLowercase{p}}\, \mathbf{\MakeLowercase{p}}$ is always a scalar.  So we have
$\ell=2$, which means that we know $\adjoint_{\mathbf{\MakeLowercase{p}}}^{3} \mathbf{\MakeLowercase{q}}$ is a
scalar multiple of $\adjoint_{\mathbf{\MakeLowercase{p}}}^{1} \mathbf{\MakeLowercase{q}}$, and so on.  This is
how Eq.~\eqref{eq:adjoint-powers} can be so simple.

Doran \etal~\cite{Doran1993} presented a particularly useful
formulation in which the universal enveloping algebra is a bivector
algebra---a subalgebra of a Clifford algebra, also known as a
geometric algebra.  The associative product is the Clifford product.
It is also known~\cite{Hestenes1987, Doran2010} that any bivector in a
space of three dimensions or fewer may be written as a blade---the
Clifford product of two anticommuting vectors.  Let us write
$\mathbf{\MakeLowercase{p}} = \bm{v}\, \bm{w}$ for some vectors
$\bm{v}$ and $\bm{w}$.  Then we have
$\mathbf{\MakeLowercase{p}}\, \mathbf{\MakeLowercase{p}} = \bm{v}\, \bm{w}\,
\bm{v}\, \bm{w} = -\bm{v}\, \bm{v}\, \bm{w}\,
\bm{w}$.  In any Clifford algebra, the product of a vector with
itself is a scalar, by definition, so the last expression is a scalar.
Thus, for any problem in which the generators are restricted to a
three-dimensional subspace, we will always obtain
Eq.~\eqref{eq:adjoint-powers}, which will always result in an
expression like Eq.~\eqref{eq:omega_logarithms}---with an
appropriately generalized definition of the angular velocity, and the
sine and cosine possibly replaced by their hyperbolic counterparts.
Slightly more generally, we do not need $\mathbf{\MakeLowercase{p}}^\ell$ to be
a \emph{scalar}, but our result is obtained as long as that term
commutes with the lower-order adjoints.  In Clifford algebra, such
conditions will frequently occur, for example, when
$\mathbf{\MakeLowercase{p}}^\ell$ is a scalar plus pseudoscalar (a generalized
complex number), which will always commute with a bivector.

The bivector presentation of, for example, Doran
\etal~\cite{Doran1993} suggests that the number of degrees of freedom
for generators in an $n$-dimensional Lie algebra should scale as
$\binom{n}{2}$, whereas the equivalent of the rotor presentation of
the group should scale as $2^{n-1}$, meaning that integration by
generators always reduces the number of equations needed---and the
benefit increases rapidly with the dimension of the algebra.  The
message is simply that we may expect cases in which the problem
simplifies, and the sum in Eq.~\eqref{eq:conjugate_adjoint} may be
evaluated exactly; the need for truncation should not be assumed.  But
with or without simplification, integration by generators may provide
an accurate and efficient approach to integrating within general Lie
groups.


\section{Numerical example}
\label{sec:numerical-examples}

To evaluate the methods presented in
Secs.~\ref{sec:integr-frame-vectors}, \ref{sec:integr-spinors},
and~\ref{sec:integr-bivectors}, we need a problem that is adequately
complicated to provide a realistic challenge, yet also simple enough
to obtain an analytical solution for comparison to the numerical
results.  First, a broad framework for developing this type of problem
is presented, along with the appropriate method of measuring the error
when different approaches may be taken.  Then, because this paper is
motivated by the precessing black-hole binary system, we will
construct a problem that emulates the types of rotation and timescales
seen in that system.  This problem will then be solved numerically and
compared to the exact answer as a means of evaluating the various
integration methods.

\subsection{Framework}
\label{sec:framework}

The discussion around Eq.~\eqref{eq:integr-const-ang-vel} showed that
it is simple to integrate rotation about a fixed axis.  It is also a
simple matter to differentiate a rotor describing rotation about a
fixed axis:
\begin{equation}
  \label{eq:differentiate-fixed-axis}
  \mathbf{\MakeUppercase{R}}(t) = \ensuremath{\mathrm{e}}^{f(t)\, \mathbf{\MakeLowercase{v}}} %
  \quad \implies \quad %
  \dot{\mathbf{\MakeUppercase{R}}}(t) = \dot{f}(t)\, \mathbf{\MakeLowercase{v}}\, \mathbf{\MakeUppercase{R}}(t)
  = \dot{f}(t)\, \mathbf{\MakeUppercase{R}}(t)\, \mathbf{\MakeLowercase{v}},
\end{equation}
for an arbitrary differentiable function $f(t)$ and an arbitrary
constant $\mathbf{\MakeLowercase{v}}$.  However, we also have the product rule
\begin{equation}
  \label{eq:product-rule}
  \mathbf{\MakeUppercase{R}}(t) = \mathbf{\MakeUppercase{R}}_{1}(t)\, \mathbf{\MakeUppercase{R}}_{2}(t) %
  \ \implies \ %
  \dot{\mathbf{\MakeUppercase{R}}}(t) = \dot{\mathbf{\MakeUppercase{R}}}_{1}(t)\,\mathbf{\MakeUppercase{R}}_{2}(t) + \mathbf{\MakeUppercase{R}}_{1}(t)\,\dot{\mathbf{\MakeUppercase{R}}}_{2}(t).
\end{equation}
This, of course, can be iterated to differentiate arbitrary products
of rotors.  We also know how to differentiate inverse rotors, by
Eq.~\eqref{eq:differentiate-unit-condition}.  But we can construct
highly nontrivial rotations just be composing them in this way, even
when each individual rotation is a rotation about a fixed axis.  For
example, define
\begin{subequations}
  \label{eq:general-product-example}
  \begin{gather}
    \mathbf{\MakeUppercase{R}}_{1}(t) = \ensuremath{\mathrm{e}}^{f(t)\, \mathbf{\MakeLowercase{v}}}, \\
    \mathbf{\MakeUppercase{R}}_{2}(t) = \ensuremath{\mathrm{e}}^{g(t)\, \mathbf{\MakeLowercase{w}}}, \\
    \mathbf{\MakeUppercase{R}}(t) = \mathbf{\MakeUppercase{R}}_{1}(t)\, \mathbf{\MakeUppercase{R}}_{2}(t)\, \mathbf{\MakeUppercase{R}}_{1}^{-1}(t).
  \end{gather}
  Here, $\mathbf{\MakeUppercase{R}}(t)$ is a rotation through the angle $2\, g(t)$ about the
  axis
  $\mathbf{\MakeLowercase{w}}'(t) = \mathbf{\MakeUppercase{R}}_{1}(t)\, \mathbf{\MakeLowercase{w}}\,
  \mathbf{\MakeUppercase{R}}_{1}^{-1}(t)$.  That is, the axis of rotation is itself rotated by
  $\mathbf{\MakeUppercase{R}}_{1}(t)$, so this is now a much more complicated rotation.  Yet
  it is very simple to compute the derivative
  \begin{equation}
    \label{eq:general-product-derivative}
    \dot{\mathbf{\MakeUppercase{R}}} = \dot{f}\, \mathbf{\MakeLowercase{v}}\, \mathbf{\MakeUppercase{R}}
    + \dot{g}\, \mathbf{\MakeUppercase{R}}_{1}\, \mathbf{\MakeUppercase{R}}_{2}\, \mathbf{\MakeLowercase{w}}\, \mathbf{\MakeUppercase{R}}_{1}^{-1}
    - \dot{f}\, \mathbf{\MakeUppercase{R}}\, \mathbf{\MakeLowercase{v}}.
  \end{equation}
\end{subequations}
It should be noted that $f(t)$ and $g(t)$ can be quite complicated,
but as long as their derivatives are given in closed form, the
derivative of $\mathbf{\MakeUppercase{R}}(t)$ can also be given in closed form.  Moreover, we
can compose several such rotations to apply different physically
motivated effects.  Section~\ref{sec:prec-nutat-binary} will show that
it is possible to emulate various features of the very complex motion
of a precessing black-hole binary using just a few simple rotations
which can easily be differentiated analytically.

Now, given an exact rotor function of time and its derivative, we can
use Eq.~\eqref{eq:angular-velocity} to find its angular velocity.  We
then apply the methods of the previous sections to integrate that
angular velocity, and compare the result to the analytical rotation to
evaluate the accuracy and efficiency of our methods.  But before we
can compare the results, we need to decide on what quantities to
compare because the methods give different types of results.  The
vector method gives a pair of vectors, the rotor method a rotor, and
the generator method a generator.  We could, for example, exponentiate
the result of the generator method to find the corresponding rotor.
But then it is not clear how to define the rotor corresponding to a
pair of vectors; it is possible, but there is ambiguity that may hide
error somehow.

Ultimately, the purpose behind integrating the angular velocity is to
be able to relate vectors in the rotating system to vectors in the
inertial system.  It is sufficient to work with bases of the inertial
and rotating systems, so we can construct an error measure that is
both simple---because it deals only with the bases---and captures all
possible error.  To do so, we define the inertial frame
$\bm{f}_{i}$, where $i$ labels the usual $x$, $y$, and $z$
directions.  Now, given an exact rotation $\mathbf{\MakeUppercase{R}}_{\text{e}}(t)$, we can define
the exact rotated frame
$\bm{e}_{i}(t) \coloneqq \mathbf{\MakeUppercase{R}}_{\text{e}}(t)\, \bm{f}_{i}\,
\mathbf{\MakeUppercase{R}}_{\text{e}}^{-1}(t)$.  On the other hand, by integrating the angular
velocity, we obtain an approximate frame $\bm{a}_{i}(t)$ using any
of the methods described in Secs.~\ref{sec:integr-frame-vectors},
\ref{sec:integr-spinors}, and~\ref{sec:integr-bivectors}.  We then
define the error norm to be
\begin{equation}
  \label{eq:error_norm}
  \delta(t) \coloneqq \sqrt{\sum_{i} \left\| \bm{e}_{i}(t) -
      \bm{a}_{i}(t) \right\|^{2}},
\end{equation}
where the norm of each vector difference is given by the usual inner
product.

To integrate the angular velocities, we use standard numerical
integration schemes.  In particular, since we are most concerned with
high-accuracy results and relatively smooth problems, we will use an
eighth-order Dormand-Prince method~\cite{hairer1993, Press2007} and
the Bulirsch-Stoer method \cite{stoer2002, Press2007}.  Because our
example problems (as well as our motivating example of binary
black-hole systems) involve smooth data and the problems are not
stiff, these integrators are the best standard general-purpose
choices.  The Bulirsch-Stoer (B-S) approach is typically capable of
taking far fewer steps than the Dormand-Prince (D-P) approach to
achieve a given accuracy.  However, B-S involves substantially more
overhead in each step, and so will frequently take more time than D-P
in these examples---even though the latter may typically take ten
times as many steps.  Nonetheless, we will find generally good
behavior with both integrators, showing that general-purpose
integrators may typically provide good results with this type of
problem.

Both integrators accept as input certain tolerances: a relative
tolerance $\text{tol}_{\text{rel}}$ and an absolute tolerance $\text{tol}_{\text{abs}}$.  If $y_{i}$
denotes each of the $N$ evolved variables, and $\delta y_{i}$ the
corresponding estimated error, then the step size of the integrator is
adjusted so that
\begin{displaymath}
  \frac{1}{N} \sum_{i} \frac{\delta y_{i}^{2}} {(\text{tol}_{\text{abs}} + \left\lvert{y_{i}}\right\rvert
    \text{tol}_{\text{rel}})^{2}}
\end{displaymath}
is less than $1$.  In each of the three approaches to integrating
angular velocity we use, the evolved variables represent components of
geometric vectors.  Those components may oscillate through $0$, and
error in each of the components should be treated the
same---regardless of the instantaneous magnitude of the variable.
Thus, we set $\text{tol}_{\text{rel}}$ to $0$ in every case, and rely only on $\text{tol}_{\text{abs}}$.
It must be noted, however, that this is error tolerance is imposed at
each step of the integration.  The total instantaneous error
$\delta(t)$ will generally grow in time with the number of steps
taken.

\subsection{Precessing and nutating binary}
\label{sec:prec-nutat-binary}
This example emulates the motion of a precessing black-hole binary
system.  Such binaries are expected to be possible sources for
gravitational-wave telescopes~\cite{Kalogera2000, OShaughnessy2005,
  Grandclement2004, Abadie2010}, occurring when one or both black
holes have significant spin components that are not aligned with the
orbital angular momentum.  The full equations that need to be evolved
to describe such a system are Einstein's equations, which constitute
an enormously complicated system of partial differential
equations~\cite{Wald1984, Misner1973}.  No exact closed-form solution
of a binary system is known, but inexact solutions can be found using
a ``post-Newtonian'' approximation~\cite{Blanchet2006a, Buonanno2003,
  Boyle2014}, in which the black holes are treated as point sources
tracked by their coordinate trajectories.  The key variables evolved
in this approach are the directions of the spins on the individual
black holes, the orbital frequency, and the orientation of the
binary.\footnote{The black-hole separation can be calculated from the
  orbital frequency, and so is not evolved as a separate variable.}
Because the differential equations resulting from the post-Newtonian
approximation are still very complicated and---most
importantly---coupled, we do not have exact solutions to them.
However, we can emulate the characteristics seen in typical precessing
evolutions in order to evaluate the best approach to integrating the
angular velocity to find the orientation of the binary.

The dominant motion is the orbit, in which the black holes simply
rotate about their common center of mass at frequency $\Omega_\text{orb}$.
We will model this rotation by the rotor
$\mathbf{\MakeUppercase{R}}_{1} = \ensuremath{\mathrm{e}}^{\Omega_\text{orb}\, t\, \mathbf{\MakeLowercase{z}} /2}$.  Exchange of angular momentum
between the orbit and the black-hole spins leads to a precession of
the orbital axis at frequency $\Omega_\text{prec}$, so that the axis roughly
traces out a cone of opening angle $\alpha$.  However, as angular
momentum is radiated in the form of gravitational waves, the orbital
angular momentum decreases, which gradually widens this precession
cone at some rate $\dot{\alpha}$, which we take to be constant.
Tilting the orbital axis down onto this cone can be achieved by the
rotor $\mathbf{\MakeUppercase{R}}_{2} = \ensuremath{\mathrm{e}}^{(\alpha + \dot{\alpha} t)\, \mathbf{\MakeLowercase{x}}/2}$, and the
precession along this cone can be achieved by rotating \emph{this}
rotor by $\mathbf{\MakeUppercase{R}}_{3} = \ensuremath{\mathrm{e}}^{\Omega_\text{prec}\, t\, \mathbf{\MakeLowercase{z}} /2}$.  Due to off-axis
components of the moment-of-inertia tensor, smaller nutations of the
orbital axis occur at the orbital frequency on an angular scale
$\nu$.  The basic tilt can be given by
$\mathbf{\MakeUppercase{R}}_{4} = \ensuremath{\mathrm{e}}^{\nu\, \mathbf{\MakeLowercase{x}}/2}$, but this rotor should also be
rotated---in this case by $\mathbf{\MakeUppercase{R}}_{1}$.  Finally, we also wish to rotate
the entire system by some rotor $\mathbf{\MakeUppercase{R}}_{0}$.

\begin{figure}
  \includegraphics[width=0.675\linewidth]{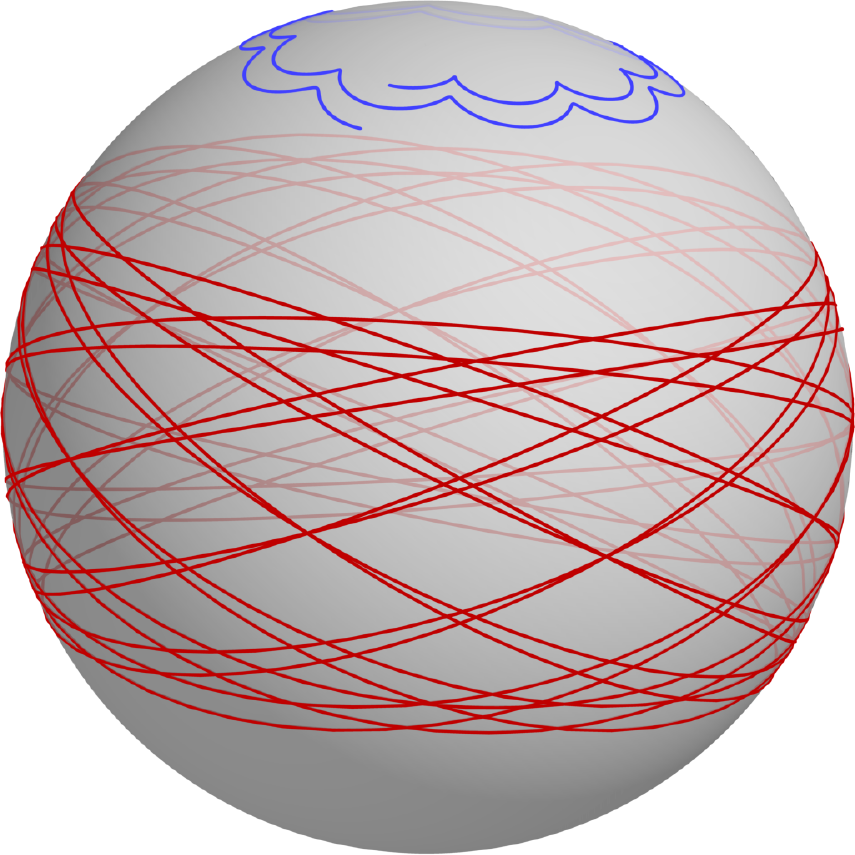}
  \caption{ \label{fig:orbital_evolution} %
    \textbf{Orbital motion in the precessing and nutating binary}.
    This figure shows the paths of the unit vectors during the first
    twenty orbits and two precession cycles (\num{20000} time units)
    in the precessing example of Sec.~\ref{sec:prec-nutat-binary}.
    The blue curve at the top shows the path traced out by the orbital
    axis (the rotated $\bm{z}$ vector).  The red curve around the
    center shows the path traced out by the unit separation vector
    between the two black holes (the rotated $\bm{x}$ vector).
    The features seen here include the fast orbital motion, visible in
    the extensive motion of the red curve; the precession, visible as
    the broadly circular motion of the blue curve; widening of the
    precession, visible as the gradually increasing radius of the blue
    curve; and nutation, visible as the scalloped shape of the blue
    curve.  These are qualitatively the same as features found in real
    precessing binary black-hole systems, but are approximated here as
    simple functions so that we have the analytical solution to
    compare to. %
  }
\end{figure}
The total rotation at any instant is then given by
\begin{equation}
  \label{eq:BinaryRotor}
  \mathbf{\MakeUppercase{R}} = \mathbf{\MakeUppercase{R}}_{0}\, \mathbf{\MakeUppercase{R}}_{1}\, \mathbf{\MakeUppercase{R}}_{4}\, \mathbf{\MakeUppercase{R}}_{1}^{-1}\, \mathbf{\MakeUppercase{R}}_{3}\, \mathbf{\MakeUppercase{R}}_{2}\,
  \mathbf{\MakeUppercase{R}}_{3}^{-1}\, \mathbf{\MakeUppercase{R}}_{1}.
\end{equation}
We choose the constants in these definitions to be comparable to
typical values during the last few orbits of a typical comparable-mass
binary with strong precession:
\begin{subequations}
  \label{eq:precession_constants}
  \begin{align}
    \Omega_\text{orb} &= 2\pi / \num{1000}, \\
    \Omega_\text{prec} &= 2\pi / \num{10000}, \\
    \alpha &= \pi / \num{8}, \\
    \dot{\alpha} &= 2\alpha / \num{100000}, \\
    \nu &= \pi / \num{80}, \\
    \mathbf{\MakeUppercase{R}}_{0} &= \ensuremath{\mathrm{e}}^{-3\alpha\, \mathbf{\MakeLowercase{x}}/10}.
  \end{align}
\end{subequations}
We will evolve this system for a total time of $\num{1000000}$
units,\footnote{All quantities in physical black-hole binaries scale
  with some power of the total mass of the system $M$.  Thus, a binary
  is generally evolved in arbitrary units; any system with the same
  mass-ratio and spin parameters is then known in physical units by
  scaling that result with $M$.  For this reason, time is typically
  measured in units of the ``geometrized mass''
  $G_{\text{N}}\, M / c^{3}$, where $G_{\text{N}}$ is Newton's
  gravitational constant and $c$ is the speed of light.  For our
  example, this means that the unit of time is irrelevant; it could be
  milliseconds or hours and---in principle---describe a physically
  possible binary.} so that the binary goes through \num{1000} orbits,
with \num{100} precession cycles, and its precession cone opens to
three times its original angle.  The evolution of a real black-hole
binary is obviously much more complicated, but these time scales
should provide a more rigorous test of the integration methods than
will typically be encountered in simulations of real systems.  The
orbital motion is depicted in Fig.~\ref{fig:orbital_evolution}, where
all of the features described above can be seen.

Now, since each of the individual rotors $\mathbf{\MakeUppercase{R}}_{1}$ through $\mathbf{\MakeUppercase{R}}_{4}$ is
a simple rotation about a constant axis, we can easily differentiate
each with respect to time.  Furthermore, we can use the product rule
to differentiate the product given in Eq.~\eqref{eq:BinaryRotor} and
obtain $\dot{\mathbf{\MakeUppercase{R}}}$, hence also $\bm{\omega}$.  Explicitly, we have
\begin{subequations}
  \begin{align}
    \label{eq:derivatives_of_rotors}
    \dot{\mathbf{\MakeUppercase{R}}}_{0} &= 0, \\ %
    \dot{\mathbf{\MakeUppercase{R}}}_{1} &= \mathbf{\MakeUppercase{R}}_{1}\, \Omega_\text{orb}\, \mathbf{\MakeLowercase{z}} /2, \\ %
    \dot{\mathbf{\MakeUppercase{R}}}_{2} &= \mathbf{\MakeUppercase{R}}_{2}\, \dot{\alpha}\, \mathbf{\MakeLowercase{x}}/2, \\ %
    \dot{\mathbf{\MakeUppercase{R}}}_{3} &= \mathbf{\MakeUppercase{R}}_{3}\, \Omega_\text{prec}\, \mathbf{\MakeLowercase{z}} /2, \\ %
    \dot{\mathbf{\MakeUppercase{R}}}_{4} &= 0.
  \end{align}
\end{subequations}
The derivatives of the inverses are found using
Eq.~\eqref{eq:differentiate-unit-condition}.  We then differentiate
Eq.~\eqref{eq:BinaryRotor} using the product rule to find $\dot{\mathbf{\MakeUppercase{R}}}$.
Then, plugging the result into Eq.~\eqref{eq:angular-velocity}, we can
determine the angular velocity analytically.  We integrate this
according to each of the methods detailed above, and finally compare
the result of the integration to the original analytical value of
Eq.~\eqref{eq:BinaryRotor}.

\begin{figure}
  \includegraphics{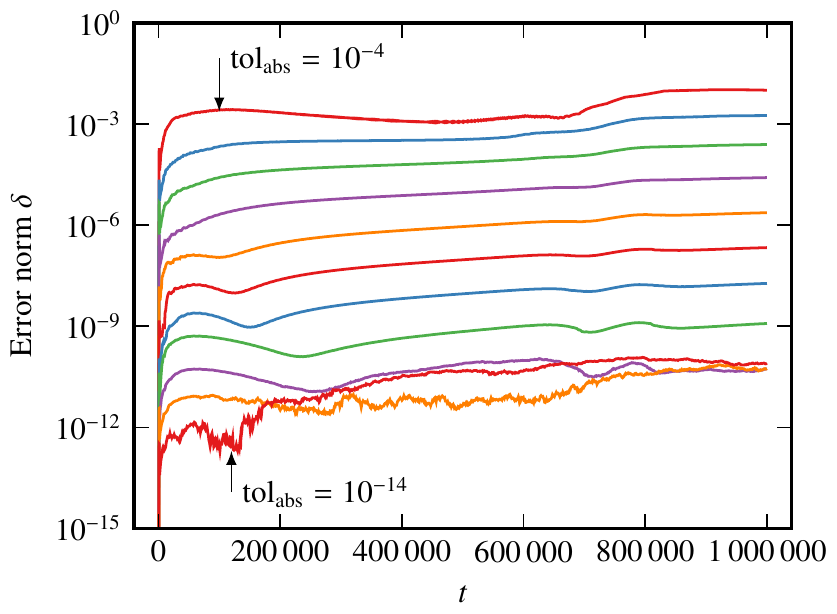}%
  \caption{ \label{fig:ErrorAsFunctionOfTime} %
    \textbf{Error norm when integrating using rotors}. This figure
    shows the error norm $\delta(t)$ given in
    Eq.~\eqref{eq:error_norm} for various choices of the absolute
    tolerance parameter, when integrating the precessing and nutating
    binary example of Sec.~\ref{sec:prec-nutat-binary} using the rotor
    approach described in Sec.~\ref{sec:integr-spinors} with the
    eighth-order Dormand-Prince integrator.
    The tolerance is decreased by a factor of $10$ for each successive
    line, demonstrating very clean convergence until the smallest
    tolerance, which seems to be limited .
    In the following figures, rather than showing the error as a
    function of time, we simply select the maximum error on each curve
    and plot this for various integration methods. %
  }
\end{figure}

As a first example, the error norm is shown for a range of tolerances
in Fig.~\ref{fig:ErrorAsFunctionOfTime}, using the Dormand-Prince
integrator to evolve the rotor.  The tolerance is decreased by a
factor of $10$ for each successive line:
$10^{-4}, 10^{-5}, \dots, 10^{-14}$.  We see the resulting error norms
also decrease by roughly a factor of $10$ each time, indicating good
convergence---except for the smallest tolerance.  As mentioned above,
the tolerance is a local tolerance imposed at each step of the
integration, so that over time the actual error should grow to a
larger value than the input tolerance, roughly proportional to the
number of steps taken.  The last two lines take on the noisy
appearance characteristic of evolutions limited by machine precision
after taking so many steps~\cite{Boyle2007a}.

\begin{figure*}
  \includegraphics{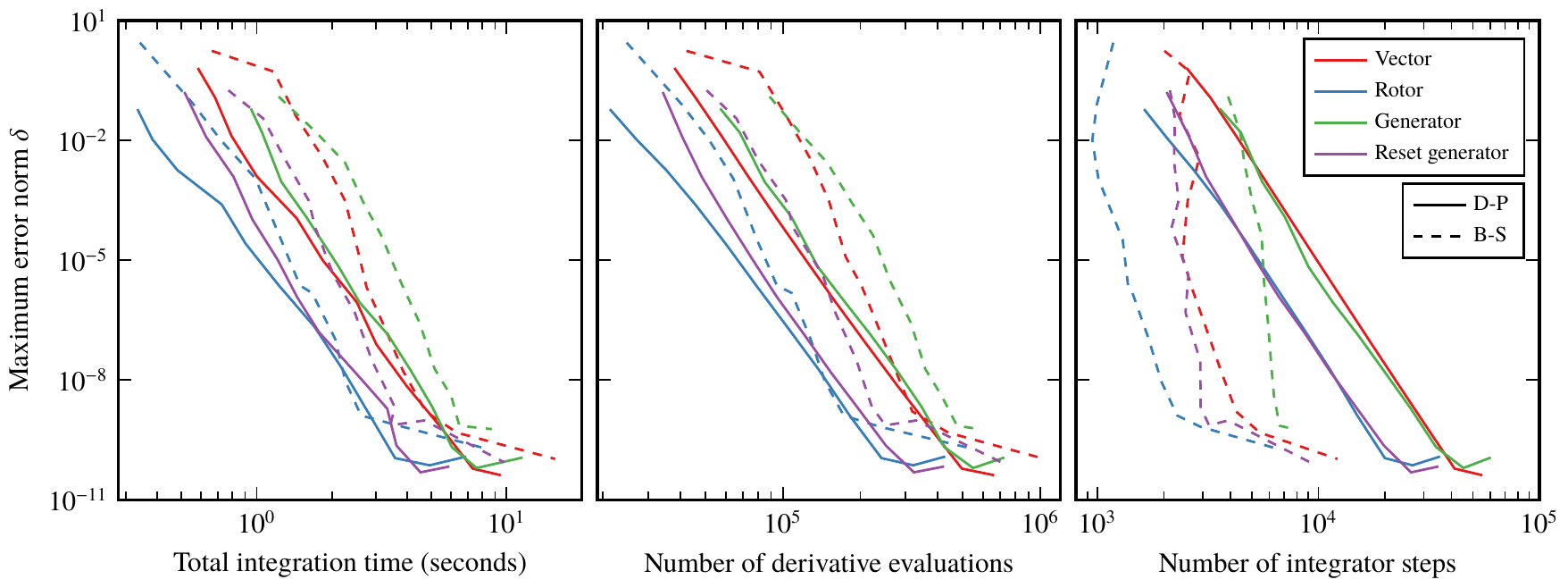} %
  \caption{ \label{fig:MaximumError} %
    \textbf{Maximum error norm for various integration methods}. This
    figure shows the maximum value of the error norm $\delta(t)$
    defined by Eq.~\eqref{eq:error_norm}, when using the various
    integration methods and numerical integrators described in the
    text.  The plot on the left shows the error as a function of the
    total (wall-clock) time taken by the integration; the center plot
    shows the error as a function of the number of evaluations of the
    derivative used to achieve that accuracy; and the plot on the
    right shows the error as a function of the number of steps taken
    by the integrator.  Along each line, different points correspond
    to different values for the absolute tolerance parameter of the
    numerical integrator as in
    Fig.~\ref{fig:ErrorAsFunctionOfTime}---typically resulting in
    longer integration times and higher accuracy for smaller
    tolerances.  %
    %
    Note that for this example, the eighth-order Dormand-Prince
    integrator (D-P; solid lines) is usually slightly faster than the
    Bulirsch-Stoer integrator (B-S; dashed lines), despite the fact
    that it requires several times as many steps at high accuracy.
    This is because the B-S integrator involves a very complicated
    algorithm.  In particular, D-P typically requires an average of
    just over \num{12} evaluations of the derivatives per step;
    whereas B-S requires anywhere from \num{20} to \num{90}, larger
    numbers being needed for smaller tolerances. %
    It is, however, notable that the B-S integrator achieves nearly
    its smallest error ($\mathord{\sim} 10^{-9}$) using just over \num{2000}
    steps, though the system goes through \num{1000} orbits during the
    evolution.  These very large steps mean that during a single step
    the B-S integrator evolves into the more rapidly varying part of
    the generator integration, even when it is reset between steps, as
    discussed below.
  }
\end{figure*}

Next, we examine the accuracy of the integration using each of the
approaches described above, as well as the Bulirsch-Stoer integrator.
Behavior like that seen in Fig.~\ref{fig:ErrorAsFunctionOfTime} is
fairly typical for these cases (as well as other test cases extracted
from Refs.~\cite{Zupan2011, zupan2014, treven2015}) but contains
somewhat more information than we need.  For clarity, the following
figures will simply take the maximum error norm for each curve, rather
than show the full dependence on time.

Figure~\ref{fig:MaximumError} shows the maximum error norm during the
integration for each of the methods described above, and for each
numerical integrator.  The plot on the left shows the error versus the
total time that integration required.  Obviously, the precise timing
will depend on the compiler, processor, and various other
details,\footnote{All computations for this paper were performed on a
  single core of an Intel Core i7 \SI{2.5}{\giga\hertz} processor.
  Except for the \texttt{lsoda} integrator (which is from the
  \texttt{scipy} package, version 0.16.1), all code was written in
  pure python (version 3.5), much of which was then automatically
  compiled as needed by the \texttt{numba} package (version 0.24),
  which uses the \texttt{LLVM} compiler (version 3.7).} but the
relative performance of the methods should be fairly consistent.
There are two major factors in the efficiency of any given approach:
the numerical integrator, and the representation of the rotation.

For this problem, the Dormand-Prince integrator typically runs
somewhat faster than the Bulirsch-Stoer (B-S) integrator, despite the
fact that---especially at high accuracies---the B-S integrator can
take far fewer steps to achieve the same errors, as seen in the plot
on the right-hand side of Fig.~\ref{fig:MaximumError}.  This is
unsurprising because the B-S algorithm is very complicated, and each
step incurs substantial overhead cost.  Its most important feature is
that it involves many evaluations of the derivatives [the right-hand
sides of Eqs.~\eqref{eq:integr-frame-vectors},
\eqref{eq:integrate-spinor}, and~\eqref{eq:log_deriv_omega}], anywhere
from \num{20} for low tolerances to \num{90} for high tolerances in
this example.  In fact, a similar plot of the total number of
evaluations of the derivatives looks almost exactly like the plot of
the total integration time, even for this example where the angular
velocity is given by a simple closed-form expression.

The second important feature of these results is the relative
efficiency of the different formulations.  For a given integrator, the
rotor formulation is always more efficient than the vector
formulation, which is always more efficient than the simple generator
formulation.  The latter point---that the generator is the least
efficient formulation---may be somewhat surprising considering that it
requires only three variables to be integrated, has no constraints
that need to be satisfied, and in the simple case of
Eq.~\eqref{eq:integr-const-ang-vel} has a simple linear solution.
However, if we reset the generator according to
Eq.~\eqref{eq:generator_reset} at the end of each time step if its
magnitude is greater than or equal to $\pi/2$, the behavior of the
generator solution is much improved, achieving efficiency that
essentially the same as that of the rotor approach with the D-P
integrator, though somewhere between that of the rotor and vector with
the B-S integrator.  We can understand these trends by looking at the
actual quantities that need to be evolved in each case.

\begin{figure}
  \includegraphics{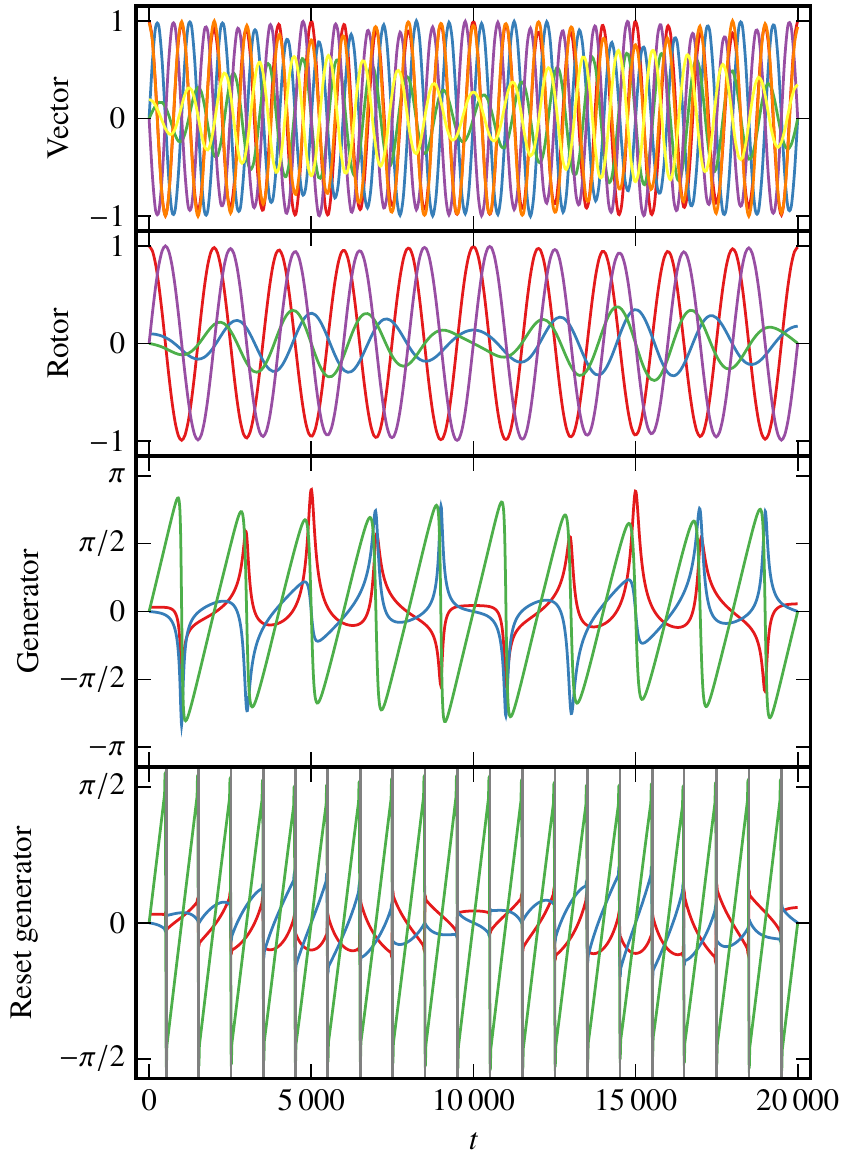} %
  \caption{ \label{fig:EvolvedQuantities} %
    \textbf{Evolved Quantities}. These plots show the actual
    quantities evolved for the vector, rotor, and generator
    formulations when solving the precessing example for the first
    $20\,000$ time units---one fiftieth of the total time.  The
    identities of the particular components do not matter; only their
    general behavior is interesting, so we do not label them.  The
    vector components vary twice as rapidly as the rotor components,
    which is a typical feature of how rotors function, due to the two
    factors of $\mathbf{\MakeUppercase{R}}$ appearing in Eq.~\eqref{eq:rotor-rotation}.  This
    suggests an explanation as to why the integrations of the vectors
    take roughly twice as long, with twice as many time steps.  More
    surprising is the behavior of the generator.  Though we frequently
    do see roughly linear behavior, each time the amplitude approaches
    $\pm \pi$, the components change very sharply.  This makes the
    system hard to integrate efficiently.  On the other hand, if we
    discontinuously reset the generator according to
    Eq.~\eqref{eq:generator_reset}, we obtain the components seen in
    the bottom panel.  The integrator does not need to evolve through
    the discontinuity, and everything in between is smooth and slowly
    varying, so integration of this quantity is much more efficient. %
  }
\end{figure}

Figure~\ref{fig:EvolvedQuantities} shows the actual quantities evolved
in each of the three systems, for the first $20\,000$ time units.  The
first point to note is that the six vector components vary twice as
quickly as the four rotor components.  This is a very general feature
of the behavior of rotors, and is due to the two factors of $\mathbf{\MakeUppercase{R}}$ found
in Eq.~\eqref{eq:rotor-rotation}; in a very rigorous sense, $\mathbf{\MakeUppercase{R}}$ is
the square-root of the usual rotation operator.  Each time the vectors
complete one cycle, the rotor completes only half a cycle.  This is
related to the spin-$1/2$ nature of rotors, and the fact that the
rotor group [which is isomorphic to $\mathrm{SU}(2)$] forms a double cover of
$\mathrm{SO}(3)$.  The slower dynamics and smaller number of components make
it entirely plausible that we should expect the rotor formulation to
be roughly twice as fast as the vector formulation in many types of
problems.  Since time step sizes are affected more directly by the
higher derivatives of the integrated functions, it is instructive to
look at the second derivative of the rotor:
\begin{equation}
  \label{eq:rotor-second-deriv}
  \frac{\ensuremath{\mathrm{d}}^{2} \mathbf{\MakeUppercase{R}}} {\ensuremath{\mathrm{d}} t^{2}} = \frac{1} {2} \dot{\bm{\omega}}\,
  \mathbf{\MakeUppercase{R}} + \frac{1} {4} \bm{\omega}^{2}\,
  \mathbf{\MakeUppercase{R}}.
\end{equation}
Now, if $2 \left\lvert{ \dot{\bm{\omega}}}\right\rvert \gtrsim \bm{\omega}^{2}$,
we can expect the rotor method to require integration steps small
enough to resolve the time dependence of $\bm{\omega}$, which will
be faster than that of $\mathbf{\MakeUppercase{R}}$.  But this will be the same in the vector
approach.  So in the worst case, we can expect time step sizes to be
comparable in the rotor and vector approaches.  We might distinguish
between vibrations, in which the system oscillates on small angular
scales with rapid time dependence, from rotations in which the angular
velocity varies relatively slowly.  Then the rotor method will lose
the factor of \num{2} advantage in vibrations.  In fact we will see
two such examples in the Appendix.  But even then, fewer equations
need to be integrated using rotors, so that there should always be at
least a small advantage.  Thus, we conclude that the rotor approach
will always be preferable to the vector approach.

The generator components vary at roughly the same frequency as the
rotor components, and we do indeed see the expected approximately
linear behavior for large portions of the evolution.  However, these
portions are punctuated by very rapid changes in the components.
These changes are caused by the system passing close to---but not
precisely through---the identity.\footnote{This is essentially the
  same as the branch-cut discontinuity familiar from the complex
  logarithm, which may be ``unwrapped'' to give a smooth curve.  The
  effect seen here is precisely a three-dimensional version of that
  discontinuity.  In the two-dimensional case, the system is
  topologically forced to return to the identity.  In three dimensions
  there is no such requirement, so the system will more typically just
  miss $\pm \pi$, and the logarithm cannot be made smoother by
  unwrapping.} %
To resolve these features adequately, the numerical integration must
take many small steps around them, leading to the poor behavior seen
in Fig.~\ref{fig:MaximumError}.  Moreover, the sharp features are
highly sensitive to the precise orientation of the system.  A slightly
different value for $\mathbf{\MakeUppercase{R}}_{0}$ in Eq.~\eqref{eq:BinaryRotor} leads to
very different behavior: much sharper or smoother curves.  Whenever
the system happens to wander close to the identity, the features will
become extremely sharp.

We can discern a pattern in these sharp features: they only occur when
the magnitude of the generator becomes large, approaching $\pi$.  As
discussed near the end of Sec.~\ref{sec:integr-bivectors}, we can
reset the generator to decrease its magnitude whenever it grows beyond
$\pi/2$.  The reset generator is shown in the bottom panel of
Fig.~\ref{fig:EvolvedQuantities}.  While there are true
discontinuities at roughly the orbital period, these are located at
discrete times; in between, the components of the generator are very
smooth.  We can apply this to numerical integration by simply imposing
the reset at the end of each step the integrator takes.  The
discontinuities do not need to be resolved in any way by the
integrator, so that they do not affect the size of the time steps it
can take.  Thus, by applying this reset, the generator approach goes
from being the slowest one seen here to being competitive with the
rotor approach.  One interesting effect of this is that the B-S
integrator can take so few steps, and hence such large steps, that
from beginning to end of the step the system may go well past the
point where it could have been reset.  That is, within a single step
the system will evolve to a very dynamical state.  And since the
generator can only reasonably be reset \emph{between} steps, this
diminishes the performance of the B-S integrator when applied to the
generator approach, which is why we see the rotor approach being
substantially more efficient.

Not shown are the results for the constraint-damped system where
$\bm{f}_{0}$ is evolved by Eq.~\eqref{eq:vector-norm-damping}, and
$\bm{f}_{1}$ is evolved by
Eq.~\eqref{eq:vector-orthogonality-damping}.  Whenever the damping
parameters $\lambda$ and $\mu$ are large enough to noticeably impact
the results, this system is orders of magnitude slower than any other
system because it is stiff.  For good measure, a third numerical
integrator was also used for this system: the \texttt{lsoda}
integrator as implemented in the \texttt{scipy}
package~\cite{scipy}, which is designed for stiff systems.  While that
does slightly improve the efficiency over the D-P and B-S integrators
for most values of $\text{tol}_{\text{abs}}$, it still cannot compete with the
efficiency of the non-damped vector system.  There may exist
applications for which such damping could be effective---perhaps when
lower-order integration is used with noisy data.  But given the
results of this example, it seems likely that the rotor or generator
approaches would be more effective in every case.

\section{Conclusions}
\label{sec:Conclusions}

This paper has presented three fundamental methods of integrating
angular velocity, along with various possible improvements.  These
methods were then evaluated by application to a rigorous test case
with an analytically known solution.  The results show that even
standard integration algorithms can deliver very accurate evolutions
with great efficiency.

The direct evolution of vectors by Eq.~\eqref{eq:integr-frame-vectors}
is elementary, and is equivalent to evolution of the rotation matrix
at the most naive level.  We can eliminate half of the redundancy in
this approach by evolving two basis vector, and computing the third as
the unique perpendicular unit vector completing a right-handed triple.
In our test case, we saw that this method approaches the best method
within a factor of \num{2} in efficiency.  But this factor of \num{2}
is likely to be a very general feature, due to the nature of
quaternion and generator representations of rotations, whenever the
angular velocity itself varies more slowly than the vectors it
describes.  This essentially make the vector twice as dynamic as the
quaternion and generator, requiring twice as many time steps during
integration.  Two constraint-damping terms were suggested in
Eqs.~\eqref{eq:vector-norm-damping}
and~\eqref{eq:vector-orthogonality-damping}, as a way to possibly
improve the accuracy of integration at a given time-step size, or
equivalently allow the integrator to take larger steps.  It turns out
that these terms simply make the system stiff, leaving it orders of
magnitude less efficient than the other approaches.  Generally, it
seems clear that direct integration of the vectors will not be the
preferred method for any system.

A better alternative is integration of the rotor responsible for the
rotation, by Eq.~\eqref{eq:integrate-spinor}.  This rotor is a nonzero
quaternion which acts on any vector according to
Eq.~\eqref{eq:rotor-rotation}, resulting in the rotated version of
that vector.  As discussed in Sec.~\ref{sec:integr-spinors}, the
unusual use of the inverse in Eq.~\eqref{eq:rotor-rotation} frees us
from the usual normalization constraint on the quaternion;
Eq.~\eqref{eq:integrate-spinor} is always the correct evolution
equation, regardless of the normalization of the quaternion.  This
allows the quaternion to provide the most efficient method of
integration found in this paper, despite the fact that the quaternion
uses four degrees of freedom to represent a rotation that has only
three intrinsic degrees of freedom.  In a way, that fourth degree of
freedom is hidden.


The final method we examined was direct evolution of the generator of
the rotation by Eq.~\eqref{eq:log_deriv_omega}.  Using quaternion
techniques it is a simple matter to relate a rotation to its
generator, with no need for intermediate translations to matrices or
other representations.  This explicitly requires just three degrees of
freedom, and simplistic arguments suggest that the components of the
generator should behave in a simple---nearly linear---manner.  It
turns out that such arguments are overly simplistic, and the
components actually undergo very rapid evolution during certain stages
of a typical rotation, as seen in the third panel of
Fig.~\ref{fig:EvolvedQuantities}.  However, it is possible to
discontinuously change these components after certain time steps,
reducing the magnitude of the generator, and making the evolution much
simpler.  With this improvement, integration by generator goes from
being the least efficient of these three methods to very closely
rivaling the rotor method.

Because of these additional complications (and perhaps the
transcendental function in its evolution equation), the generator
method cannot generally be expected to be as robustly efficient as the
rotor method, though different situations may provide a minor
advantage to one or the other.  If the system is restricted to very
small rotations---staying in the neighborhood of the identity---the
discontinuities will never come into play and the generator method may
be very slightly more efficient than the rotor method.  This may be
the case in twisting of beams, for example, where a complete rotation
of the beam from its original position may be uncommon.  However, for
larger rotations, the rotor method will frequently be able to achieve
a given accuracy with fewer steps because the rotors never enter the
delicate region for which the generator reset was introduced.  Taken
together, these considerations suggest that the rotor approach should
be a good general choice unless specific features of the problem
recommend the generator.

Nonetheless, the differences between the rotor and generator methods
are fairly slight.  Section~\ref{sec:note-about-munthe} describes
generalizations of the generator method to other Lie groups.  In
particular, it suggests a method for finding exact evolution equations
in many cases---as opposed to resorting to finite truncations as in
the Magnus and Munthe-Kaas algorithms.  Such a method finds possible
application in integrating relativistic motions in the Lorentz group
$\mathrm{SO}^{+}(3,1)$, or even the conformal spacetime representation
$\mathrm{Spin}(4,2)$ which can incorporate translations~\cite{Doran2010}.
These topics, of course, are beyond the scope of this paper.

A final conclusion may also be drawn from these results.  Quite
simply, the general-purpose numerical integrators used here are
capable of evolving rotations very accurately and stably when used
with an adequate formalism.  The example of the precessing black-hole
binary is rigorous, involving both large rotations due to the basic
orbital nature of the system, as well as smaller precessional and
nutational oscillations on very different timescales.  Nonetheless,
over many multiples of these dynamical timescales, the integrators are
able to achieve high accuracy, approaching the limits of machine
precision.

\begin{acknowledgments}
  It is my pleasure to thank Scott Field, Larry Kidder, and Saul
  Teukolsky for useful conversations, as well as Nils Deppe for
  particularly enlightening discussions of stiffness and numerical
  integrators.  I also appreciate Eva Zupan and Miran Saje for useful
  comments on their papers, and for pointing out more recent
  references.  This project was supported in part by the Sherman
  Fairchild Foundation, and by NSF Grants No.\ PHY-1306125 and
  AST-1333129.
\end{acknowledgments}


\begin{figure*}[!ht]
  \includegraphics{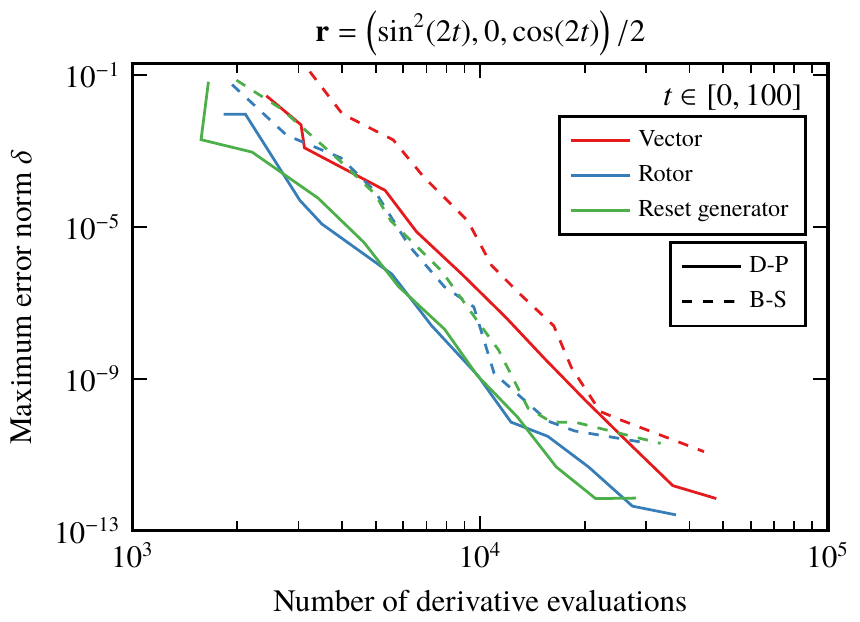}%
  \hfill%
  \includegraphics{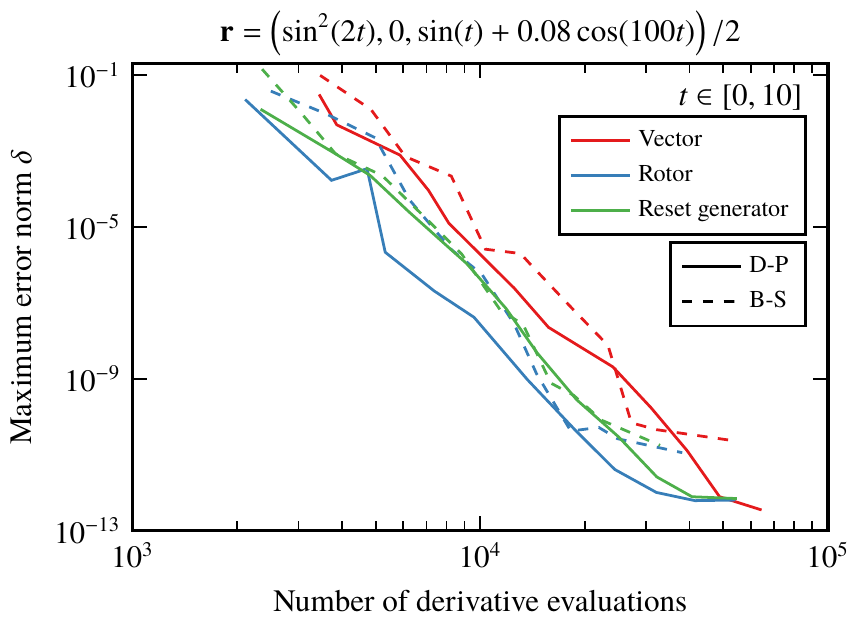}%
  \\[12pt] %
  \includegraphics{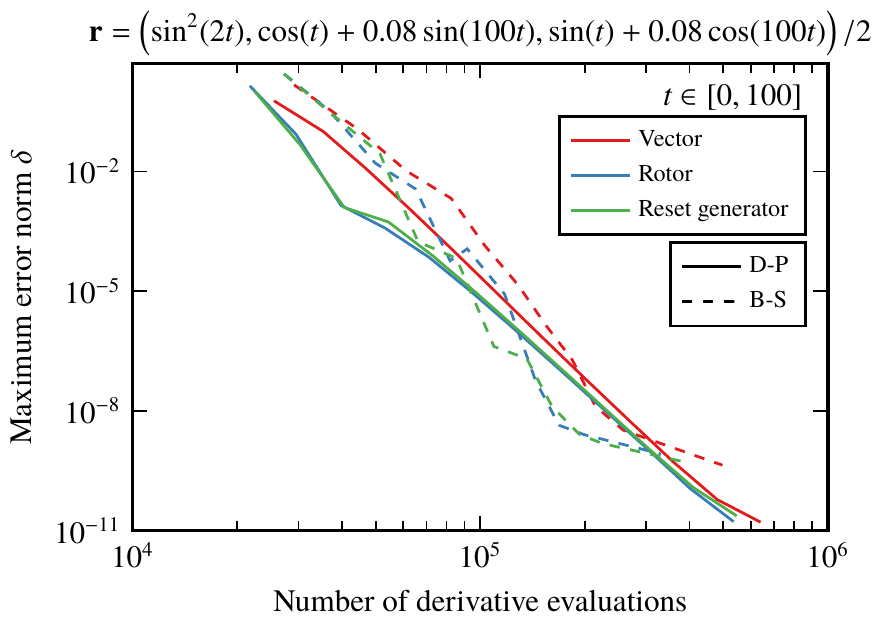}%
  \hfill%
  \includegraphics{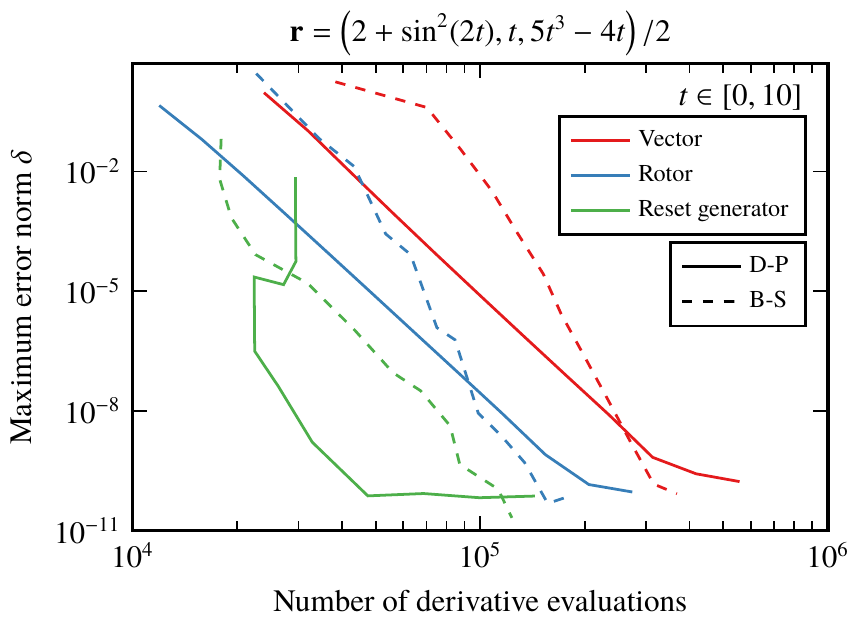}%
  \caption{ \label{fig:ZupanSaje} %
    \textbf{Numerical examples of Zupan-Saje}. These plots show the
    same quantities as in the center plot of
    Fig.~\ref{fig:MaximumError}, except that each system is one of
    the four examples of Zupan and Saje~\cite{Zupan2011}.  The exact
    generators are shown in the title of each plot, which are
    integrated over the range of times shown above the legends.  The
    data give the error of the integrated angular velocity for values
    of $\text{tol}_{\text{abs}}$ ranging from $10^{-4}$ to $10^{-14}$.
  }
\end{figure*}

\appendix* 

\section{Additional numerical examples}
\label{sec:additional-numerical-examples}

For the sake of additional and more direct comparisons to other
references, this appendix briefly presents several more numerical
examples.  Zupan and Saje~\cite{Zupan2011} constructed four examples
by inventing analytic functions to be taken as the components of the
rotation vector $\bm{\vartheta}(t)$, which is essentially the
generator of the rotation.  In the language of quaternions this is
just twice the logarithm, so
$\mathbf{\MakeLowercase{r}} = \bm{\vartheta}/2$.  Now, because the
expressions for $\bm{\vartheta}$ are given as simple functions of
time, we can also find the derivative as
$\dot{\mathbf{\MakeLowercase{r}}} = \dot{\bm{\vartheta}}/2$.  Thus, we can
obtain analytic expressions for the angular velocity $\bm{\omega}$
using Eq.~\eqref{eq:omega_logarithms}.  We can then integrate the
angular velocity to deduce the frame, and compare that result to the
analytic result of directly using the rotation operator
$\mathbf{\MakeUppercase{R}} = \ensuremath{\mathrm{e}}^{\mathbf{\MakeLowercase{r}}}$.

Plots of the maximum error norm versus the number of steps taken by
the integrator are shown in Fig.~\ref{fig:ZupanSaje} for each of the
Zupan-Saje examples.  The exact generators are listed in the
respective titles.  The upper right and lower left, in particular,
show examples of systems that might be called vibrational; their
angular velocities vary quickly relatively to the frames they
describe.  In particular, both systems satisfy
$2 \left\lvert{ \dot{\bm{\omega}}}\right\rvert \gg \bm{\omega}^{2}$, especially
the system shown in the lower left.  As discussed below
Eq.~\eqref{eq:rotor-second-deriv}, this means that the numerical
integrators must track the evolution of $\bm{\omega}$, which means
that the rotor method does not have such a large advantage over the
vector method.  Nonetheless, the rotor approach is slightly more
efficient even in these cases.

It must be noted that the approach used to devise these numerical
examples is highly synthetic, and can lead to very unrealistic
motions.  In particular, the unbounded growth of the generator in the
final example (lower right plot) exhibits extremely large and variable
rotations, yet returns precisely to the identity rotation many times
with increasing frequency.  In fact, the naive implementation of the
generator method breaks down entirely with this example, as the
equations become stiff.  The reset generator method shown here,
however, is actually the most efficient one in that case, precisely
because it is well suited to this type of rotation.  With the notable
exception of that one line, we generally obtain roughly the same
behavior seen in the example of Sec.~\ref{sec:prec-nutat-binary}: the
rotor and generator methods being comparable, and the vector method
being substantially less efficient.  Moreover, comparison to the
results found by Zupan and Saje~\cite{Zupan2011} shows that the
high-order general-purpose integrators used here provide far more
accurate results.  Note that more recent work~\cite{zupan2014,
  treven2015} by those authors and collaborators obtained improved
precision using highly specialized integrators.



\vfil


\let\c\Originalcdefinition \let\d\Originalddefinition
\let\e\Originaledefinition \let\i\Originalidefinition

\bibliography{References,References2}


\end{document}